\documentclass[useAMS,usenatbib]{PaperIII}

\usepackage{hyperref}
\usepackage{graphicx}
\usepackage{bigstrut}
\usepackage{threeparttable}
\usepackage{multirow}
\usepackage{array}
\usepackage{booktabs}
\usepackage{enumitem}
\usepackage{xcolor}
\usepackage{stfloats}
\usepackage{amssymb}
\usepackage{amsmath}

\newcommand{\lime}{{\sc{LIME}}}
\newcommand{\radmc}{{\sc{RADMC-3D}}}
\newcommand{\casa}{{\sc{CASA}}}

\newcolumntype{C}[1]{>{\centering\let\newline\\\arraybackslash\hspace{0pt}}m{#1}}

\defcitealias{Evans&Ilee2015}{Paper I}
\defcitealias{Evans&Ilee2017}{Paper II}

\graphicspath{{Figures/}{Figures/SensitivityCharts/}{Figures/LineSpectra/}{Figures/PVDiagrams/}{Figures/MassEstimates/}{Figures/Appendices/}}

\title[GIs in a protosolar-like disc III]{Gravitational instabilities in a protosolar-like disc III: molecular line detection and sensitivities}
\author[]{
M.\ G.\ Evans$^{1}$\thanks{E-mail: py09mge@leeds.ac.uk},
T.\ W.\ Hartquist$^{1}$, 
P.\ Caselli$^{2}$, 
A.\ C.\ Boley$^{3}$,
J.\ D.\ Ilee$^{4}$,
\newauthor
J.\ M.\ C.\ Rawlings$^{5}$\smallskip\\  
$^{1}$School of Physics and Astronomy, University of Leeds, Leeds LS2 9LN, UK\\
$^{2}$Max-Planck-Institut f$\ddot{u}$r extraterrestrische Physik, Giessenbachstrasse, 85741 Garching bei M$\ddot{u}$nchen, Germany\\
$^{3}$Department of Physics and Astronomy, 6224 Agricultural Road, Vancouver, BC V6T 1Z1, Canada\\
$^{4}$Institute of Astronomy, University of Cambridge, Madingley Road, Cambridge CB3 0HA, UK\\
$^{5}$Department of Physics and Astronomy, University College London, London WC1E 6BT, UK}

\begin{document}

\date{Accepted Year Month Day. Received Year Month Day; in original form 2018 Apr 25}

\maketitle

\begin{abstract}

At the time of formation, protoplanetary discs likely contain a comparable mass to their host protostars. As a result, gravitational instabilities (GIs) are expected to play a pivotal role in the early phases of disc evolution. However, as these young objects are heavily embedded, confirmation of GIs has remained elusive. Therefore, we use the radiative transfer code \lime\ to produce line images of a $0.17\,\mathrm{M}_{\odot}$ self-gravitating protosolar-like disc. We note the limitations of using \lime\ and explore methods to improve upon the default gridding. We synthesise noiseless observations to determine the sensitivities required to detect the spiral flux, and find that the line flux distribution does not necessarily correlate to the abundance density distribution; hence performing radiative transfer calculations is imperative. Moreover, the spiral features are seen in absorption, due to the GI-heated midplane and high extinction, which could be indicative of GI activity. If a small beamsize and appropriate molecular line are used then spatially resolving spirals in a protosolar-like disc should be possible with ALMA for an on-source time of 30\,hr. Spectrally resolving non-axisymmetric structure takes only a tenth as long for a reasonable noise level, but attributing this structure to GI-induced activity would be tentative. Finally, we find that identifying finger-like features in PV diagrams of nearly edge-on discs, which are a direct indicator of spirals, is feasible with an on-source time of 19\,hr, and hence likely offers the most promising means of confirming GI-driven spiral structure in young, embedded protoplanetary discs.

\end{abstract}

\begin{keywords}
stars: pre-main-sequence, stars: circumstellar matter, protoplanetary
discs, submillimetre: stars, planetary systems
\end{keywords}

\section{Introduction}

\begin{figure*}
    \centering
    \includegraphics[width=0.85\textwidth]{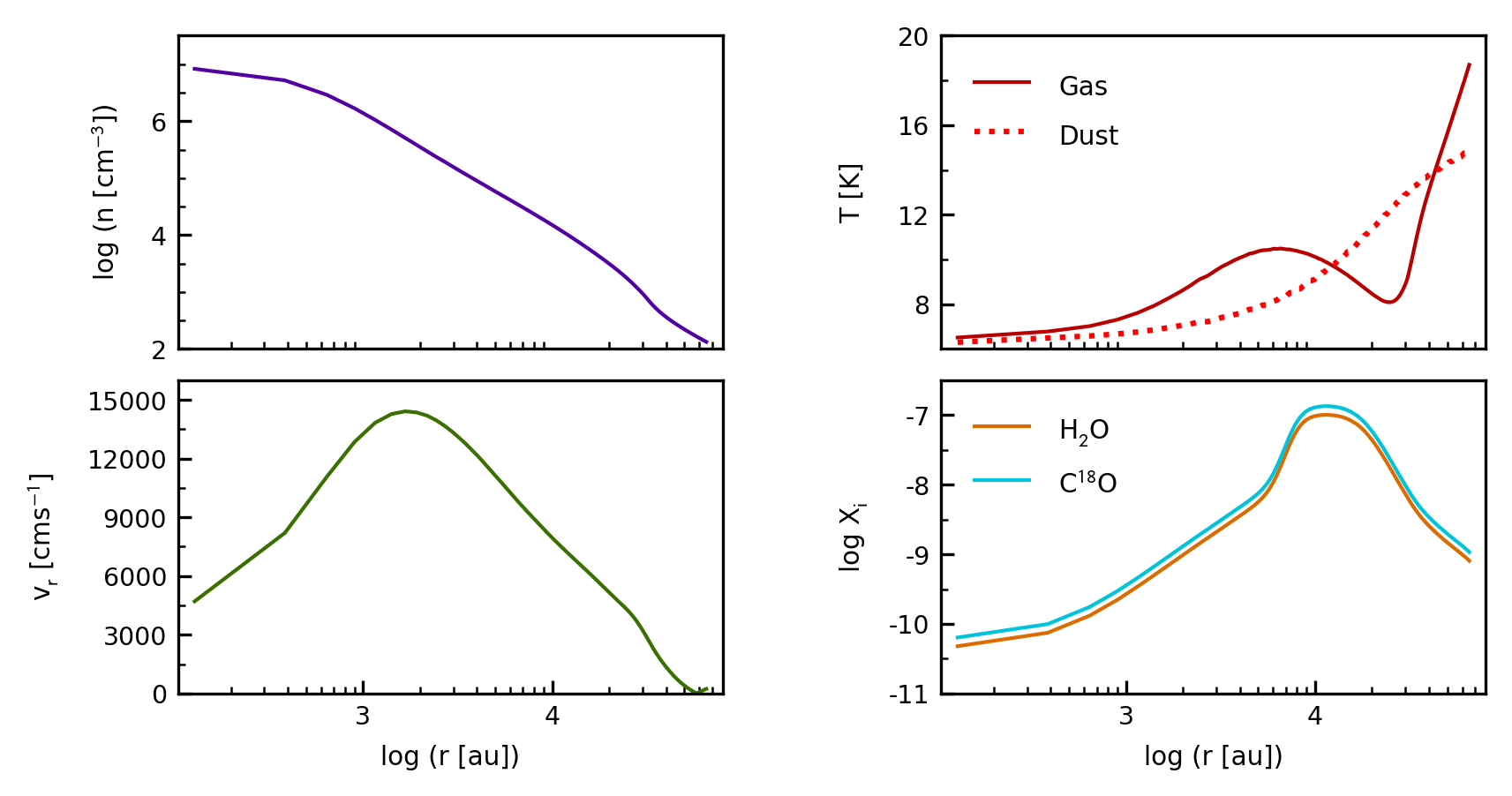}
    \caption{Spherically symmetrical envelope model used in the radiative transfer calculations \citep{Keto&Caselli2015}.}
    \label{fig:envelopepars}
\end{figure*}

Young, embedded protoplanetary discs can have a mass similar to or greater than that of their host protostar at the epoch of protostellar formation \citep[see e.g.][]{Machida&Matsumoto2011, Gerin&Pety2017}. In such situations, the disc's self-gravity can play an important role in the evolution of the system if the stabilising effects of shear and sound waves are overcome by the growth of gravitational modes. This balance is formalised by the Toomre parameter, Q, such that $Q$ = $c_s \kappa$/$\pi G \Sigma$, where $c_s$, $\kappa$, $G$, and $\Sigma$ are the sound speed, the epicyclic frequency, the gravitational constant, and the mass surface density, respectively. If $Q$ $\lesssim$ 1.7, then a smooth disc with small density perturbations is susceptible to the growth of spiral asymmetries \citep{Durisen&Boss2007} due to the gravitational instability \citep{Gammie2001, Lodato&Rice2004}. These spiral waves produce shocks as the system evolves into a non-linear regime \citep[see e.g.][]{Harker&Desch2002, Boley&Durisen2008, Bae&Hartmann2014}, and these shocks have a marked impact on the temperature of disc material locally. This can affect the chemical evolution of some molecular species as shown in \citet{Ilee&Caselli2011} and \citet{Evans&Ilee2015}, hereafter \citetalias{Evans&Ilee2015}. Moreover, this temperature structure results in a large flux contrast between spiral and inter-spiral regions in the (sub)millimetre regime. Simulations have shown that this flux contrast can translate into observable spiral structure in the millimetre continuum \citep[][hereafter \citetalias{Evans&Ilee2017}]{Cossins&Lodato2010, Douglas&Caselli2013, Dipierro&Lodato2014, Hall&Forgan2016, Evans&Ilee2017}, and recent observations of young discs may support these findings \citep[e.g.][]{Perez&Carpenter2016, Tobin&Kratter2016}.

\smallskip

As millimetre continuum observations of young, protoplanetary discs trace the dust emission exclusively, the information that can be extracted from continuum images is limited. Therefore many observers probe frequencies of specific molecular transitions in order to probe the bulk of the gas mass, which allows for a more thorough understanding of both disc morphology and composition. The inventory of molecules detected in more evolved (Class II) protoplanetary discs at (sub)millimetre wavelengths includes CO, CN, CS, C$_2$H, CH$^+$, HCN, HNC, H$_2$O, HCO$^+$, NH$_3$, N$_2$H$^+$, N$_2$D$^+$, SO, DCO$^+$, DCN, H$_2$CO, c-C$_3$H$_2$ and some isotopologues of the listed species \citep[e.g.][]{Dutrey&Guilloteau1997, Thi&Menard2011, Dutrey&Wakelam2011, Chapillon&Guilloteau2012, Qi&Oberg2013, Guilloteau&Reboussin2016, Salinas&Hogerheijde2016, Salinas&Hogerheijde2017}. ALMA is now expanding this catalogue to include more complex organic molecules such as CH$_3$CN and CH$_3$OH \citep[e.g.][]{Oberg&Guzman2015, Walsh&Loomis2016}.

\smallskip

Whilst the majority of molecular line detections have been made in more evolved discs (Class II/III), detections of CO (J = 2$\,\rightarrow\,$1), $^{13}$CO (J = 2$\,\rightarrow\,$1), C$^{18}$O (J = 2$\,\rightarrow\,$1), H$_2$CO (J = 3$_{0,3}\rightarrow2_{0,2}$), SO (J$_N$ = 6$_5\rightarrow5_4$), CS (J = 5$\,\rightarrow\,$4), CCH (N = 3$\,\rightarrow\,$2, J = 5/2$\,\rightarrow\,$3/2, F = 3$\,\rightarrow\,$2), CH$_3$OH (J$_\mathrm{N}$ = 4$_{0,4}\rightarrow3_{−1,3}$, J$_\mathrm{N}$ = 7$_{-1,7}\rightarrow6_{−1,6}$) have been made in younger (Class 0/I) systems \citep{Tobin&Hartmann2012, Tobin&Kratter2016, Ohashi&Saigo2014, Sakai&Oya2014, Sakai&Oya2016}. However, in terms of the focus of this paper, it is debatable whether the objects that have been observed thus far are gravitationally unstable. Moreover, the expected width of GI-driven spirals is on the order of au scales, but the smallest angular resolutions attained in these observations are 0.36 $\times$ 0.25\,arcsec$^2$ for L1448 IRS3B at a distance of 230\,pc \citep{Tobin&Kratter2016} and 0.65 $\times$ 0.37\,arcsec$^2$ for IRAS 04365+2535 at a distance of 140\,pc \citep{Sakai&Oya2016}, which correspond to physical scales of 80 $\times$ 60\,au$^2$ and 90 $\times$ 50\,au$^2$, respectively. Hence, the beamsizes of observations reported to date are too large to resolve any spiral structure if it exists. It is therefore very timely and prudent to investigate whether spiral structure can be resolved with higher angular resolutions, as the results can then be used to inform the selection of future observational targets. This then allows the best observational strategy to be developed for determining if spiral features within young protoplanetary discs are GI-driven.

\smallskip

In Section \ref{sec:limeimages} we explore methods to improve the default behaviour of the Line Modelling Engine \citep[\textsc{lime};][]{Brinch&Hogerheijde2010} radiative transfer code, implementing some of the optimisations found in \citetalias{Evans&Ilee2017} as a starting point. We also compare LTE and non-LTE calculations and provide insight into which frequencies should be adopted for observations of embedded protoplanetary discs. In Section \ref{sec:obs} we produce perfect, noise-free synthetic observations of our disc model across a large parameter space of molecular transitions, disc inclinations and angular resolutions. From this we produce charts that depict the sensitivity required to detect flux originating from the spiral features. We then investigate how feasible it is to spatially resolve the spiral structure from integrated intensity maps, and how feasible it is to spectrally resolve non-axisymmetric structure from line spectra and position-velocity (PV) diagrams. Finally in Section  \ref{sec:conclusions} we present our conclusions.

\section{Producing line image maps with \lime}
\label{sec:limeimages}

\subsection{Disc model}

The disc model used in this paper is the same as used in \citetalias{Evans&Ilee2017}, which is also described in detail in \citetalias{Evans&Ilee2015}. As a brief overview, the disc is gravitationally unstable and has a mass of 0.17M$_\odot$ and a radius of approximately 50\,au. It is evolved over 2050\,yr using a 3D radiative-hydrodynamics code, which includes cooling via radiative energy losses and heating via $PdV$ work, shock heating and irradiation by the central protostar. The radiative transfer treatment consists of flux-limited diffusion radially, and raytracing vertically. A gas-to-dust mass ratio of 100:1 is assumed, and the dust grains are assumed to be thermally coupled and well mixed with the gas since significant dust settling is not expected in such turbulent systems; in our disc model the Stokes number computed for mm and cm dust grains is only greater than unity at the very edges ($r$ \textgreater\ 50\,au), though we note that \citet{Dipierro&Pinilla2015} have shown that dust enhancement can still occur within spiral arms at these dust grain sizes. \citet{DAlessio&Calvet2006} dust opacities are adopted in the hydrodynamic simulation and, for self-consistency, also in the radiative transfer calculations performed in Section \ref{sec:numpoints} and thereafter. Regarding the radiative transfer, we use dust temperatures from the hydrodynamic simulation, rather than self-consistently calculating them in the radiative transfer code, in order to correctly account for the shock heating in our GI-driven disc model.

\smallskip

An envelope is not included in the radiative hydrodynamic simulation as we are assuming the disc is well shielded, i.e. heating from external sources is neglected. In terms of the effects on chemistry, accretion from the envelope can alter the disc composition, and hence synthesised line observations, but observations have indicated that variations are likely to only be significant near the centrifugal barrier \citep[e.g.][]{Sakai&Sakai2014, Sakai&Oya2017}. Outflows are also known to harbour distinct chemical signatures, but their effects on the disc and envelope are likely to be insignificant. Indeed, \citet{Drozdovskaya&Walsh2015} have shown that wide cavities are needed to directly irradiate the envelope and affect the composition. The interactions between a disc and its envelope may also affect the development of spiral structure, in particular by favouring the excitation of lower $m$ modes \citep[e.g.][]{Harsono&Alexander2011}. However, as the spectrum could be affected by a host of conditions, the overall influence of an envelope on a particular disc is difficult to predict. As such, the current simulation still represents a plausible disc response to gravitational instability and is sufficient for the purposes of a basic investigation of the detectability of GI-induced structure.

\smallskip

We do, however, account for the contribution of an envelope in the \lime\ radiative transfer calculations by implementing a 10\,M$\odot$ Bonnor-Ebert sphere \citep[see][]{Keto&Caselli2015}. The properties of this envelope model are shown in \autoref{fig:envelopepars} and we note that, whilst the velocities at the innermost radii are perhaps lower than expected in a young star-forming system, the rotational velocity is likely to be larger than the infall velocity at small radii \citep[see e.g.][]{Ohashi&Saigo2014}. Consequently, the effect of underestimated envelope velocities should be minimal provided the disc and envelope flux can be separated.

\subsection{Determining frequencies}
\label{sec:determiningfreqs}

\begin{figure*}
        \includegraphics[width=0.85\textwidth]{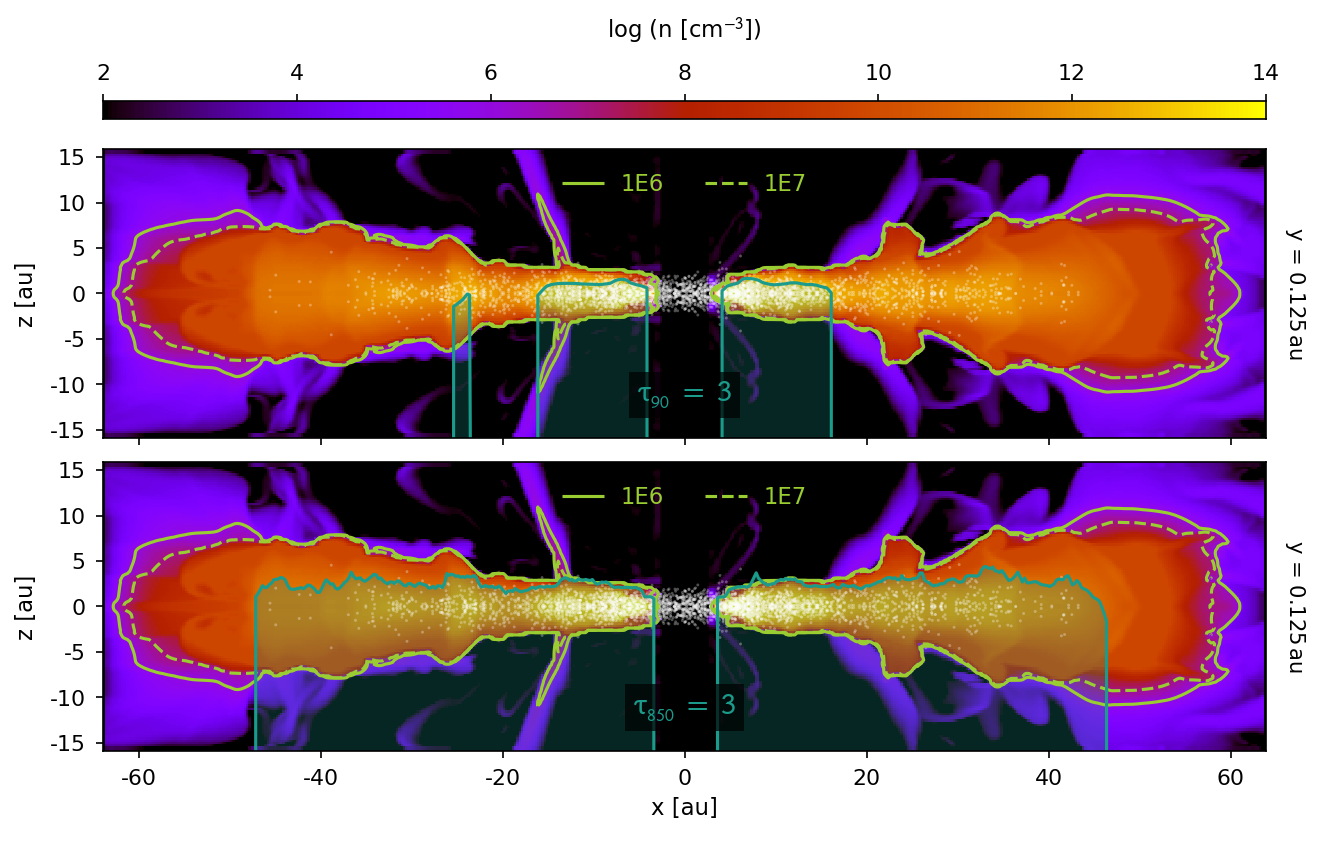}
    \caption{Nuclei number densities of the disc model along the $y$ = 0.125\,au plane with solid and dashed green lines overplotted at number densities that approximately demarcate the boundaries of LTE validity. The teal lines show the boundary where $\tau$ = 3 at a frequency of 90\,GHz (top) and 850\,GHz (bottom); hence the shaded regions indicate the optically thick regions of the disc when viewed from above. The white dots indicate the $x$ and $z$ positions of all of the fluid elements (i.e. not only those constrained to $y$ = 0.125\,au) that are chemically processed (see \citetalias{Evans&Ilee2015}).}
    \label{fig:tau3surfaceslice90850GHz}
\end{figure*}

\autoref{fig:tau3surfaceslice90850GHz} shows the surface at which the optical depth in the disc is $\tau$ = 3 at 90\,GHz (top panel) and 850\,GHz (bottom panel), superposed on y-axis slices of the number density. At 850\,GHz, the disc is largely optically thick, which means observations will only trace the surface layers of the disc. At 90\,GHz however, which is near the lowest frequency ALMA is capable of observing, the disc is less optically thick and more of the disc material is probed. This has implications for extracting reliable information from the observations. Hence we opt to produce line intensity maps at low frequencies (i.e. Band 3 of ALMA).

\smallskip

Whilst observing at low frequencies is necessary to trace the bulk of the disc material, it is also pivotal that abundance information is available in optically thin regions. In the disc model we use here the density and temperature are interpolated from the 3D hydrodynamic grid to 2000 fluid elements at intervals throughout the simulation, and then the abundances of a selection of species are calculated for each fluid element using a chemical model (see \citetalias{Evans&Ilee2015} for details).

\smallskip

\autoref{fig:tau3surfaceslice90850GHz} shows the distribution of the fluid elements within our disc model, in white, collapsed to 2D along the $y$ axis, and as can be seen, the fluid elements have a much coarser resolution than the underlying hydrodynamic grid. Consequently, at 850\,GHz, because the $\tau$ = 3 boundary is near the surface of the disc, we are only observing regions devoid of fluid elements, and therefore devoid of abundances. At 90\,GHz the disc is optically thinner as mentioned earlier, and so observations will sample more abundance-rich regions and therefore be more reliable. Note that even at low frequencies the observable regions of the innermost disc are still sparsely populated by fluid elements, and thus results for $r$ \textless\ 15\,au are perhaps questionable. However, our primary focus is on the spiral features in the outer disc and hence this is not a large concern. We would like to emphasise here that \autoref{fig:tau3surfaceslice90850GHz} also demonstrates how important it is for chemical simulations of discs to ensure the resolution of abundances is sufficient in optically thin regions, so that reliable synthesised images can be obtained.

\subsection{LTE approximation}

The complexity of line radiative transfer can be drastically simplified if local thermodynamic equilibrium (LTE) can be assumed. In this case, instead of iterating towards a converged level population distribution, the level populations can be assumed to follow a Maxwell-Boltzmann distribution. The condition for LTE is $n_{\rm{c}}^{\ k}$\ \textgreater\ $n_{\rm{crit}}^{\ \ \ k}$, where $n_{\rm{c}}^{\ k}$ and $n_{\rm{crit}}^{\ \ \ k}$ are the number density and critical density for the collisional partner $k$. The critical density is defined as the ratio of radiative to collisional transition, i.e.

\begin{equation}
    n_{\rm{crit}}^{\ \ \ k} = \frac{A_{\rm{ul}} + B_{\rm{ul}}\bar{J} - \frac{n_{\rm{l}}}{n_{\rm{u}}}B_{\rm{lu}}}{\sum_{i \neq u}\gamma_{\rm{ui}}^{\ \ k}}\,,
\end{equation}
where $n_{\rm{u}}$ and $n_l$ are the level populations of the upper and lower levels and $\gamma_{\rm{ui}}$ is the collisional rates of all transitions from the upper energy level. $A_{\rm{ul}}, B_{\rm{ul}}$ and $B_{\rm{lu}}$ are the Einstein coefficients for spontaneous emission, stimulated emission and stimulated absorption, respectively, and $\bar{J}$ is the mean intensity. 

\smallskip

We can assess whether LTE is appropriate by comparing Table 1 of \citet{Shirley2015} with the number density of our disc model, under the assumption that H$_2$ is the dominant collisional partner for all transitions. \autoref{fig:tau3surfaceslice90850GHz} shows the number density in the $y$ = 0.125\,au plane with contours of density overplotted. We have used $n_{\rm{crit}}$ for comparison purposes because it presents a worst-case scenario for LTE validity, and because $n_{\rm{eff}}$ can be undefined if an integrated line intensity of 1\,km\,s$^{-1}$ is not observed. As can be seen, the number density of the disc model exceeds the optically thin critical density in the emitting regions (i.e. $\tau$ \textless\ 3), meaning LTE should be appropriate for our disc model across a wide range of molecular transitions. We later verify this assumption in Section \ref{sec:nonlte} by comparing the LTE results to full non-LTE calculations.

\subsection{Weighting of grid points}

In \citetalias{Evans&Ilee2017} we showed that default number density weighting parameters are inadequate for our disc model. Here, we utilise abundance density weighting because our focus is on the line emission; hence we amend the weighting function to use abundance density, with a cutoff threshold of 0.001 $\times$ max(X$_i$), as only the strongest emission regions are of interest.

\subsection{Number of grid points}
\label{sec:numpoints}

In \citetalias{Evans&Ilee2017} we demonstrated that the number of grid points has a significant effect on the accuracy of the simulated continuum flux image. Line radiative transfer calculations add a level of complexity, even if LTE is assumed, so we expect more grid points than we used in \citetalias{Evans&Ilee2017} to be necessary. In order to investigate this, we produce two C$^{17}$O 1$\,\rightarrow\,$0 images ($i$ = 30$^{\circ}$) using different numbers of grid points and compare the residual integrated line fluxes. As the computational cost of producing line images is much larger than the computational cost of producing continuum images, we limit the maximum number of grid points in this paper to 1 $\times$ 10$^6$, as opposed to the 2 $\times$ 10$^6$ grid points used in the `canonical' image of \citetalias{Evans&Ilee2017}.

\begin{figure*}
        \includegraphics[width=0.95\textwidth]{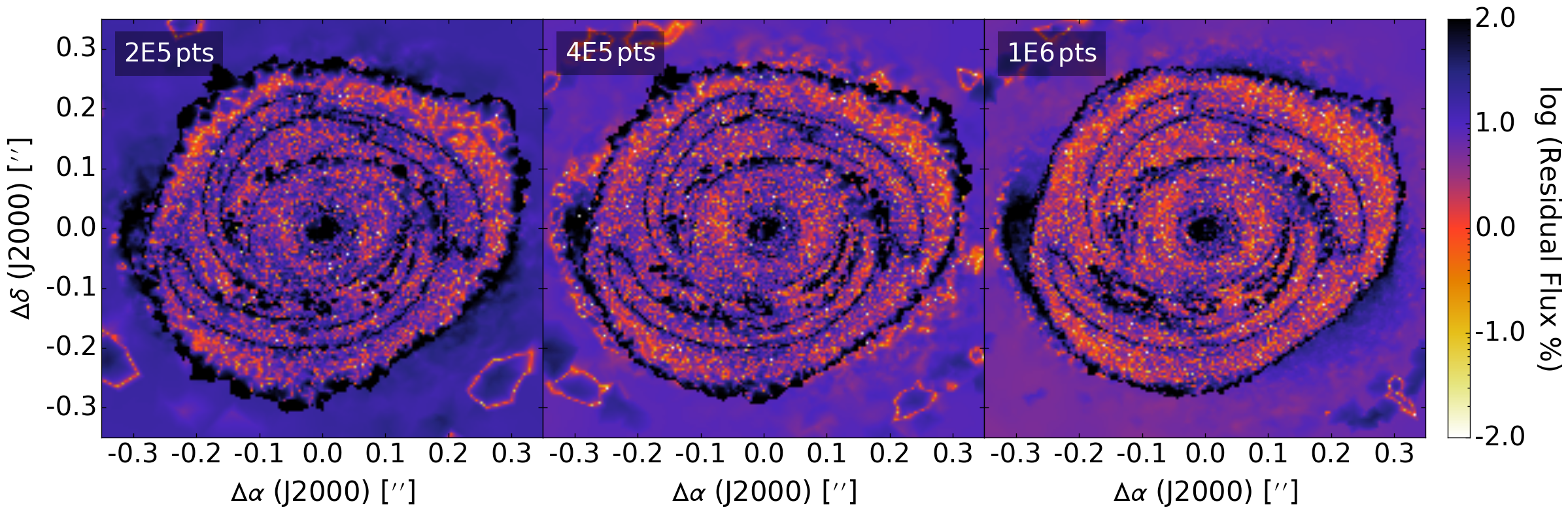}
    \caption{Residual integrated line flux for C$^{17}$O 1$\,\rightarrow\,$0 calculated under the assumption of LTE, between two \lime\ runs using the indicated number of grid points, with standard sampling. We define this residual as $|$(f$_1$-f$_2$)$|$/(f$_1$+f$_2$), where $f$ denotes integrated flux per pixel.}
    \label{fig:pointscomparison}
\end{figure*}

\autoref{fig:pointscomparison} demonstrates that as the number of grid points is increased, the integrated flux maps become more consistent, suggesting that the line images are more accurate. This is expected, but, because of the complexity of line transfer, even when LTE is assumed, the residual at 1 $\times$ 10$^6$ points is much larger than for the same number of points in the continuum \citepalias[see][Figure 5]{Evans&Ilee2017}. Therefore, in order to achieve a residual flux that is less than the expected observational errors ($\geq$ 10 per cent), more than 1 $\times$ 10$^6$ points are needed for our disc model when default sampling is used. Note that the large residual differences defining the outline of the spirals are caused by particularly large density gradients, which we discuss further in Section \ref{sec:nonlte}.

\subsection{Sampling optimisations}

Constructing a grid of more than 1 $\times$ 10$^6$ points is computationally expensive and time consuming. Consequently, it becomes advisable to investigate whether we can obtain more accurate images without increasing the number of grid points.

\subsubsection{Model-specific sampling}
\label{sec:sampling3}

The default behaviour of \lime\ is to implement either a random Cartesian sampling or a radially logarithmic sampling routine across the entire model. As of \lime\ v1.7 there is also the ability to use a completely separate tree-building sampling algorithm, but here we focus on the original `randoms-via-rejection' algorithm. So far, including in \citetalias{Evans&Ilee2017}, we have used radially logarithmic sampling, which we refer to as `Sampling2', to ensure that, probabilistically, more points are positioned within the disc than the envelope. Consequently, however, more points are positioned in the innermost disc rather than the outer disc where the spiral features of particular interest are located. In order to improve upon this behaviour, we develop a new sampling routine that uses rectilinear sampling within the disc in order to ensure a more even distribution of points. We continue to populate the envelope using radially logarithmic sampling, but only with a small fraction of the total number of points because the mean free path in the envelope is much larger than in the disc \citepalias[see][Section 2.3]{Evans&Ilee2017}. We refer to this new sampling algorithm as `Sampling3'.

\begin{figure*}
        \includegraphics[width=0.95\textwidth]{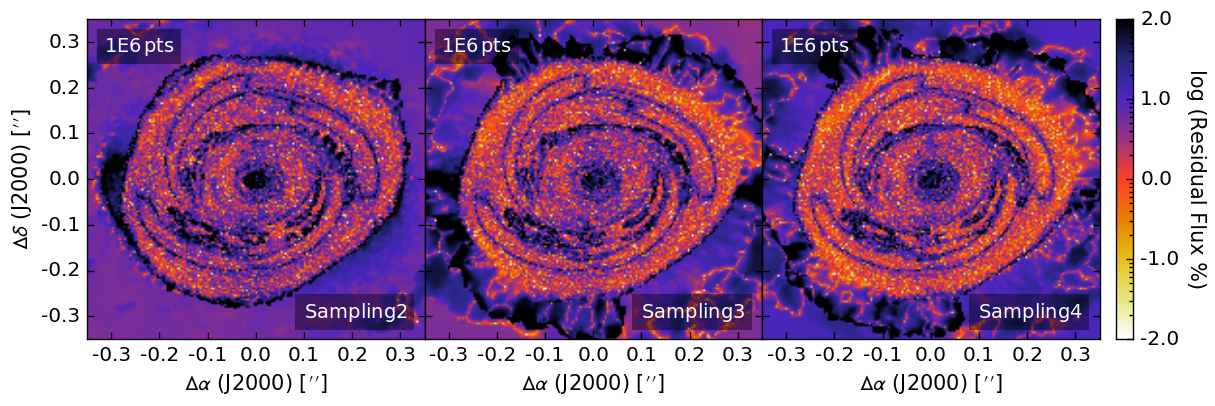}
    \caption{Residual integrated line flux for C$^{17}$O 1$\,\rightarrow\,$0 calculated under the assumption of LTE, between two \lime\ runs using the indicated number of grid points, with standard sampling (left), a model-specific sampling routine (middle), and a model-specific sampling routine with an optical depth surface to constrain the positioning of grid points (right). We define this residual as $|$(f$_1$-f$_2$)$|$/(f$_1$+f$_2$), where $f$ denotes integrated flux per pixel.}
    \label{fig:samplingcomparison}
\end{figure*}

\begin{figure*}
        \includegraphics[width=0.95\textwidth]{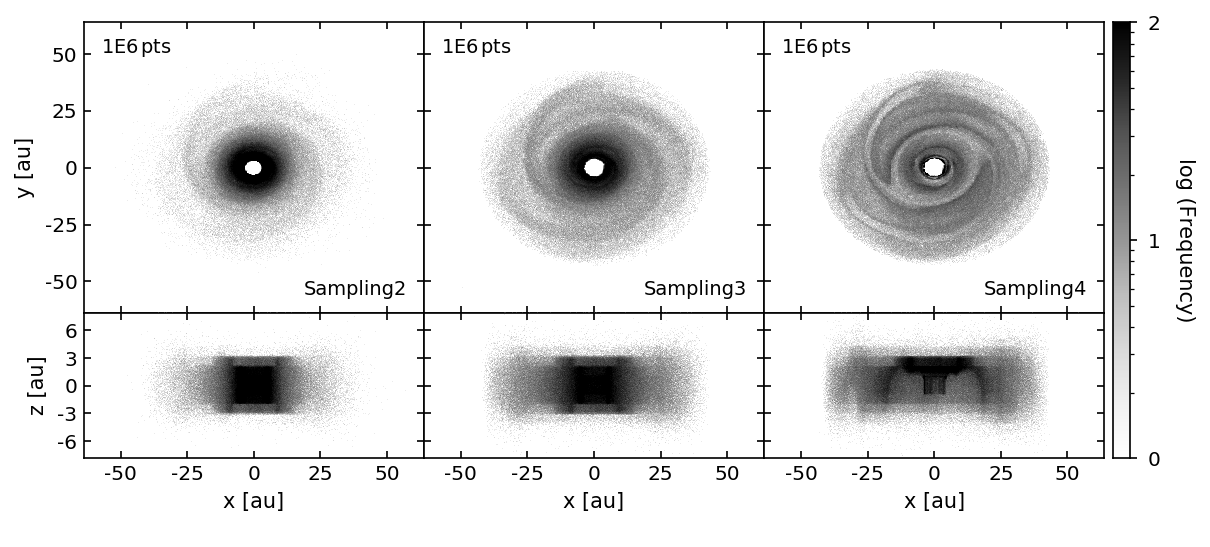}
    \caption{2D histogram of grid point positions in the $x$-$y$ plane (top) and $x$-$z$ plane (bottom) when using standard sampling (left), a model-specific sampling routine (middle), and a model-specific sampling routine with an optical depth surface to constrain the positioning of grid points (right). The disc is being viewed from above.}
    \label{fig:gridhistogram}
\end{figure*}

The middle panel of \autoref{fig:samplingcomparison} shows the residual for C$^{17}$O 1$\,\rightarrow\,$0 ($i$ = 30$^{\circ}$) when 1 $\times$ 10$^6$ grid points are used and Sampling3 is adopted. When compared to Sampling2, one of the default \lime\ algorithms, the residual within the outer disc region is improved considerably. This is because, as mentioned earlier, Sampling3 affords a much more even distribution of grid points; hence a larger proportion of grid points are positioned beyond the inner disc, as \autoref{fig:gridhistogram} demonstrates. Within the inner disc ($r$ \textless\ 15\,au) the difference in residual is minimal. This is because the natural consequence of a greater number of grid points in the outer disc is a smaller number of grid points in the inner disc. However, because the inner disc is largely optically thick, emission from points near the midplane will be negligible. Therefore the grid point positioning can be optimised further by omitting negligibly-emitting regions. Again, we note that the large residual differences defining the outline of the spirals are caused by particularly large density gradients, and discuss this further in Section \ref{sec:nonlte}.

\subsubsection{Optical depth surface}
\label{sec:optdepth}

In \citetalias{Evans&Ilee2017} we explored the notion of optimising grid point placement by restricting points to regions where $\tau$ \textless\ 3, and found this to be a viable method. This is because the contribution to intensity is diminished by optically thick material and therefore at $\tau$ \textgreater\ 3 this contribution is negligible. As the continuum absorbs radiation at a particular frequency independent of its source, we should be able to safely assume the same logic for line emission.

\smallskip

We produce a new C$^{17}$O 1$\,\rightarrow\,$0 image ($i$ = 30$^{\circ}$) using the configuration in Section \ref{sec:sampling3}, whilst employing the optical depth surface method described in \citetalias{Evans&Ilee2017}, with one adjustment: rather than relocate a point from $\tau$ \textgreater\ 3 to $\tau$ \textless\ 3 and then re-evaluate, we instead, for the sake of simplicity, reject the point and pick another randomly. We refer to this sampling algorithm as `Sampling4'. 

\smallskip

The right panel of \autoref{fig:samplingcomparison} shows the residual integrated line flux for Sampling4. Within the outer disc there is a small overall improvement in residual flux between Sampling3 and Sampling4, even though the $x$-$y$ plane images in \autoref{fig:gridhistogram} suggest that the coverage within the spiral arms is more sparse for Sampling4. However, this is because the positioning of the grid points is more optimal as intended. Indeed, by restricting points to $\tau$ \textless\ 3 we are omitting negligibly-emitting regions that lie at high optical depths. This also explains why there is a more marked improvement in the inner disc residual when using Sampling4, where the point restriction is visually evident in the $x$-$z$ plane images of \autoref{fig:gridhistogram}. We note that, whilst the residual beyond the disc boundary is considerably larger when using our custom sampling routines (Sampling3 and Sampling4), our focus is on the disc itself so this is not a concern for our results.

\smallskip

We conclude from this analysis that Sampling4 offers the most convergent result for our disc model, when 1 $\times$ 10$^6$ grid points are used. However, we would like to emphasise here that convergence is not necessarily synonymous with accuracy, and where possible results should be compared to a `canonical' result. In this case, we can compare our LTE runs using our custom sampling algorithm to a full non-LTE run adopting the default sampling.

\subsection{Non-LTE comparison}
\label{sec:nonlte}

When calculating level populations under non-LTE conditions, the user must ensure that enough iterations are used to reach a satisfactory level of convergence. In \lime\ there is no convergence threshold that, once surpassed, terminates the calculation of level populations. Instead the user can set the number of iterations to be performed. \lime\ does, however, calculate a signal-to-noise ratio for each grid point using the level population of the current iteration and the standard deviation of the level population mean over the past five iterations. This information could be used in future versions of \lime\ to develop a convergence criterion, e.g. iterating stops once 90 per cent of grid points have reached a SNR that exceeds 100. For now, however, we compare the SNR across iterations in order to determine how many iterations should be adopted for our disc model. As \autoref{fig:snrs} indicates, this convergence level appears to be reached by five iterations, but we use ten iterations for the non-LTE runs in order to err on the side of caution.

\smallskip

\begin{figure}
        \includegraphics[width=0.475\textwidth]{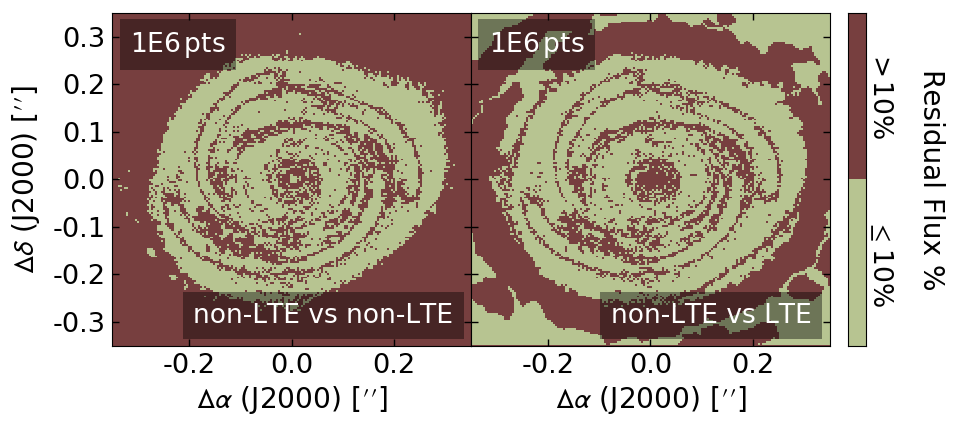}
    \caption{Residual integrated line flux for C$^{17}$O 1$\,\rightarrow\,$0 between two \lime\ runs using the indicated number of grid points, with regions exceeding 10 per cent coloured red and elsewhere coloured green. The left panel shows the result for two non-LTE runs, with standard sampling, and the right panel shows the result for one non-LTE run with standard sampling, and one LTE run with a custom sampling routine (see Section \ref{sec:optdepth}). We define this residual as $|$(f$_1$-f$_2$)$|$/(f$_1$+f$_2$), where $f$ denotes integrated flux per pixel.}
    \label{fig:ltenonltecomparison}
\end{figure}

We generate two `canonical' results by performing full non-LTE calculations for two grids with 1 $\times$ 10$^6$ points, selected with Sampling2. We opt for Sampling2 in order to avoid the possibility that our custom sampling algorithms are converging to a significantly different solution than the \lime\ default. We then calculate the residual between these two non-LTE runs, and compare to the residual of one non-LTE run and one LTE run from Section \ref{sec:optdepth}. For the non-LTE runs we ensure enough iterations are used to reach a satisfactory level of convergence for the level populations (see \autoref{sec:convergence}). To make this comparison easier to digest, in \autoref{fig:ltenonltecomparison} we demarcate regions where the residual flux is greater than 10 per cent (the expected observational errors), in red/brown, and elsewhere in green. The purpose of this comparison is to assess whether using LTE is appropriate as we have predicted, and to establish the propinquity concerning the thermodynamic configurations.

\autoref{fig:ltenonltecomparison} shows that C$^{17}$O 1$\,\rightarrow\,$0 \lime\ images produced with LTE treatment are consistent with images produced with full non-LTE treatment, which is as expected. Moreover, as \autoref{fig:tau3surfaceslice90850GHz} shows, this conclusion should hold across all low-energy transitions of the molecular species on which we focus. Therefore, this result validates our use of LTE in Section \ref{sec:obs}, which is computationally beneficial because the assumption of LTE affords a considerable speedup in level population calculations. Consequently, hereafter we use Sampling4 and assume LTE.

\smallskip

We note that within all of the residual integrated line flux maps presented thus far, there is a particularly large difference that traces the outline of the non-axisymmetric structure, regardless of whether LTE is assumed or not. This is due to the large density gradients found at the boundaries between spiral and non-spiral material. A dedicated sampling routine would need to be implemented in order to accurately account for these rapid density variations, but, as these gradients only exist in very narrow regions (on the order of the beamsize for the highest angular resolution attainable with ALMA), the overall effect on our observational results should be negligible.

\section{Synthetic observations}
\label{sec:obs}

\begin{figure*}
    \includegraphics[width=0.875\textwidth]{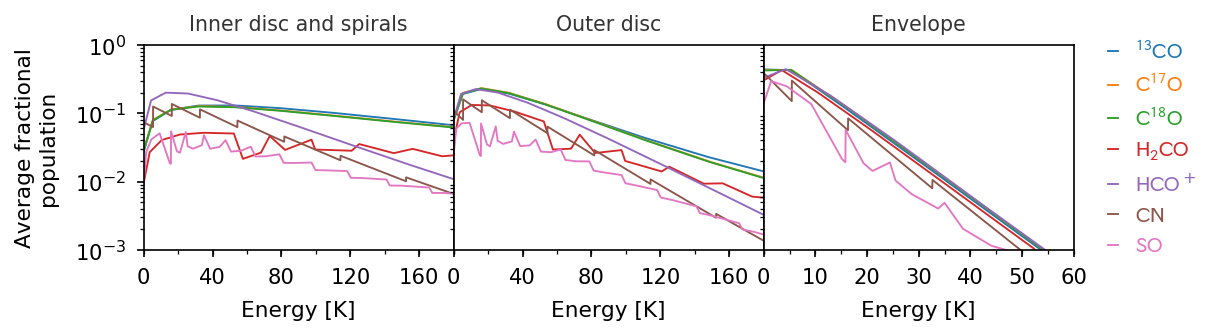}
    \caption{Fractional energy level populations averaged over grid points within the inner disc and spirals (left panel), the outer disc (middle panel) and the envelope (right panel). The disc regions are determined by density ranges.}
    \label{fig:levelpops}
\end{figure*}

We opt to produce synthetic observations of $^{13}$CO, C$^{18}$O, C$^{17}$O, H$_2$CO, HCO$^+$, CN and SO as these are species that play fundamental roles in disc chemistry and trace some aspect of the spiral structure in our disc model \citepalias[see][]{Evans&Ilee2015}. As we have shown in Section \ref{sec:determiningfreqs}, we are restricted to low frequency transitions of these molecules; hence we choose to synthesise transitions in ALMA Band 3 (84--119\,GHz), Band 6 (211--275\,GHz) and Band 7 (275--370\,GHz). In order to narrow the possible transitions down further, we use \lime\ to calculate the level populations (in LTE) of each molecule we consider and average these across different regions of our model. The results are shown in \autoref{fig:levelpops} for the inner disc and spirals, the outer disc, and the envelope.

\begin{table*}
    \centering
    \caption{Molecular transitions used to produce synthetic observations of our disc model.}
    \label{tab:transitions}
    \newcommand{\cspace}{\hspace*{6.0em}}
    \begin{tabular}{lcccccc}
        & & & & & \cspace & \cspace \\ [-2.5ex]
        \toprule
        Molecule & Transition & Upper energy [K] & Lower energy [K] & Frequency [GHz] & \multicolumn{2}{c}{Angular resolution [arcsec]} \bigstrut[b] \\
        \cline{6-7}
        & & & & & out02 & out28 \bigstrut[t] \\ 
        \midrule
        $^{13}$CO  &  J = 3$\,\rightarrow\,$2  &  22.055  &  11.028  &  330.588  &  0.993  &  0.014 \\  
        C$^{17}$O  &  J = 3$\,\rightarrow\,$2  &  22.487  &  11.244  &  337.061  &  0.974  &  0.013 \\
        C$^{18}$O  &  J = 3$\,\rightarrow\,$2  &  21.971  &  10.986  &  329.331  &  0.997  &  0.014 \\
        H$_2$CO  &  J$\rm{_N}$ = 3$_{0,3}\rightarrow2_{0,2}$  &  14.566  &  7.286  &  218.222  &  1.505  &  0.021 \\
        H$_2$CO  &  J$\rm{_N}$ = 10$_{2,8}\rightarrow11_{0,11}$  &  167.304  &  158.689  &  258.296  & 1.271  &  0.018 \\
        HCO$^+$  &  J = 3$\,\rightarrow\,$2  &  17.850  &  8.925  &  267.558  &  1.227  &  0.017 \\
        CN  &  N = 3$\,\rightarrow\,$2, J = 5/2$\,\rightarrow\,$3/2  &  22.678  &  11.335  &  340.031  &  0.966  &  0.013 \\
        CN  &  N = 2$\,\rightarrow\,$1, J = 3/2$\,\rightarrow\,$1/2  &  11.335  &  3.786  &  226.333  &  1.449  &  0.020  \\
        CN  &  N = 1$\,\rightarrow\,$0, J = 3/2$\,\rightarrow\,$1/2  &  3.786  &  0.000  &  113.495  &  2.893  &  0.040  \\
        SO  &  N = 6$\,\rightarrow\,$5, J = 7$\,\rightarrow\,$6  &  33.050  &  24.316  &  261.844  & 1.254  &  0.017  \\
        SO  &  N = 5$\,\rightarrow\,$4, J = 5$\,\rightarrow\,$4  &  30.654  &  23.475  &  215.221  &  1.526  &  0.021  \\
        SO  &  N = 2$\,\rightarrow\,$1, J = 1$\,\rightarrow\,$2  &  10.987  &  3.100  &  236.452  &  1.389  &  0.019  \\
        \bottomrule
    \end{tabular}
\end{table*}

Our focus is on the spiral features. Therefore, we identify the regions of the most populated energy levels from \autoref{fig:levelpops}, whilst also considering the fact that lower frequency transitions are more reliable for our model (see Section \ref{sec:determiningfreqs}). The outcome of this process is a list of transitions, \autoref{tab:transitions}, that we opt to produce synthetic observation for. H$_2$CO (10$_{2,8}\rightarrow$11$_{0,11}$) is included in this list because \autoref{fig:levelpops} shows that a transition between levels with relatively high energies should only be visible in the inner disc and spirals and therefore may offer a clearer picture of the innermost density structure. CN (1$_{3/2}\rightarrow0_{1/2}$) and CN (3$_{5/2}\rightarrow2_{3/2}$) are included in order to investigate the effect of transition frequency on flux; we use CN for this because the fractional populations vary the least across low energies in the inner disc and spiral regions.

\smallskip

In the envelope we use the CO profile calculated by \citet{Keto&Caselli2015} for $^{13}$CO, C$^{17}$O and C$^{18}$O, scaled by the appropriate isotopic abundances given by \citet{Wilson1999}. The HCO$^+$ profile follows the H$_2$O profile of L1544, scaled so the maximum fractional abundance is 1 $\times$ 10$^{-8}$. H$_2$CO follows a step profile with a fractional abundance of 1.5 $\times$ 10$^{-8}$ at $r$ $\leq$ 8000\,au and 1.5 $\times$ 10$^{-9}$ at $r$ \textgreater\ 8000\,au \citep{Young&Lee2004}. Finally, for SO and CN we adopt constant fractional abundances of 1 $\times$ 10$^{-10}$ \citep{Miettinen2016, Koumpia&Semenov2017}. Whilst these abundance profiles in the envelope are primitive, we expect that isolation of the envelope flux will be straightforward. Therefore, as the focus of this work is on the disc spirals, the effect on the presented results should be minimal.

\smallskip

We assume that the model replicates the same object used in \citetalias{Evans&Ilee2017}, which is positioned at a distance of 145\,pc to represent the nearest star-forming regions (e.g. the Ophiuchus Molecular Cloud). We adopt inclinations of 15$^\circ$, 30$^\circ$, 45$^\circ$, 60$^\circ$ and 75$^\circ$ in order to span the range of inclinations between the individual sources in real systems such as IRAS 16243--2422, which is a protostellar binary with nearly face-on and nearly-edge on discs separated by 600\,au. Our observations are synthesised across different fully operational ALMA antenna configurations (`out02', `out08', `out13', `out18', `out21', `out24' and `out28'), for which a selection of angular resolutions are shown in \autoref{tab:transitions}, and appropriate channel widths (see \autoref{tab:almaspecres}); we assume that each antenna configuration adopted can be used across all frequencies.

\smallskip

Perfect observations are produced with no sources of noise. This allows us to constrain the maximum attainable flux across our parameter space, which we accomplish by convolving the \lime\ images with a Gaussian beam. The motivation for this is threefold. Firstly, we can achieve a superior efficiency using a fast Fourier transform algorithm rather than producing realistic synthetic observations across our entire parameter space with software such as \casa\ \citep[v4.5.0;][]{McMullin&Waters2007}. Secondly, it is expected that a significant portion of our parameter space will simply be unobservable, meaning that producing realistic observations in this scenario is superfluous. 
Finally, because ALMA’s $u$-$v$ coverage is extensive, a Gaussian beam should be an appropriate approximation; although rather than use a circular Gaussian beam, we increase the accuracy of this approximation by using an elliptical Gaussian beam with a size determined by the \textit{simobserve} task in \casa\, for an on-source observation time of 12\,hr and Briggs weighting with a robustness factor of 0.5.

\subsection{Detecting spiral structure}

\begin{figure*}
    \centering
    \includegraphics[width=0.85\textwidth]{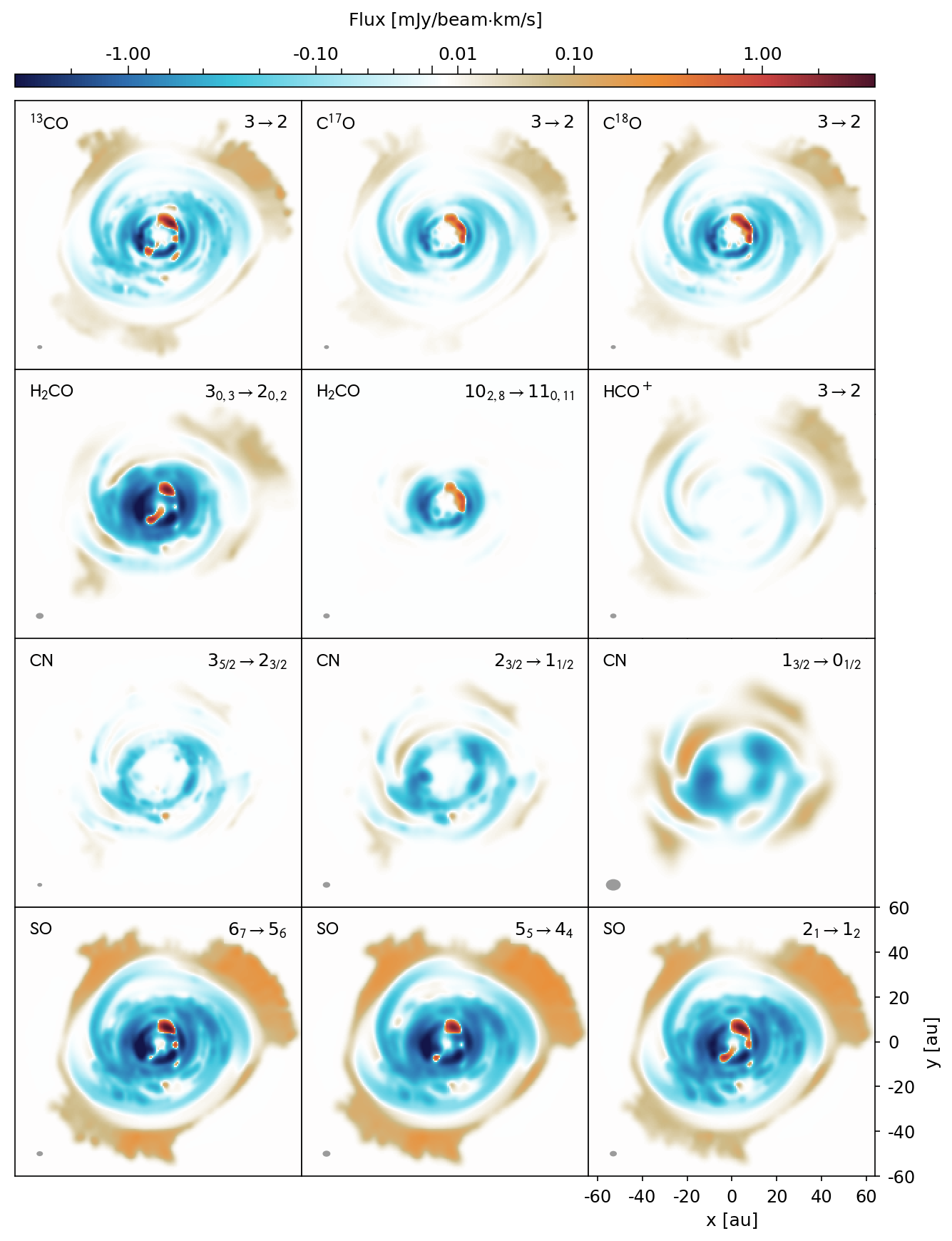}
    \caption{Noise-free integrated intensity maps (integrated from -5.0 to -0.8\,km/s and 0.8 to 5.0\,km/s) for each molecular transition, for an inclination of 15$^\circ$, a channel spacing of 31\,kHz and the `out28' antenna configuration. The bottom right panel indicates the physical scale of the observed disc model, which is constant across all transitions shown. The grey circle in each panel shows the beamsize for the corresponding observation, but note that the level of detail shown here is not currently attainable with ALMA. For comparison purposes, the corresponding column density maps can be found in \citetalias{Evans&Ilee2015}.}
    \label{fig:integratedintensitymaps}
\end{figure*}

\autoref{fig:integratedintensitymaps} shows noiseless flux maps integrated across a velocity range that avoids the bulk of the envelope contribution, for the indicated transition, narrowest channel spacing and highest angular resolution (see \autoref{tab:transitions}), at a disc inclination of 15$^\circ$. We present this figure in order to illustrate the extent of the spiral structure that is distinguishable for each species, under the condition of perfect observations, but emphasise that obtaining this level of detail is not possible with real observations.

\smallskip

\autoref{fig:integratedintensitymaps} allows us to determine the perfect observable characteristics for various molecules, which then can inform the preparation of realistic observation proposals. For example, because CO is ubiquitous due to its highly volatile nature, the spirals are distinctly traced by each isotopologue. HCO$^+$ also traces the spirals but with a weaker flux primarily due to its lower abundance density. There is an inner hole evident in the HCO$^+$ intensity map, and this results from the destruction of HCO$^+$ in the innermost disc due to high abundance of H$_2$O. The extent of this hole could potentially offer insight into the disc mass \citepalias[see][]{Evans&Ilee2015}. H$_2$CO is mostly confined to the innermost disc due to its relatively high binding energy to dust grains, and hence could offer detailed views of the inner disc structure if the angular resolution is sufficient. However, H$_2$CO also traces the spiral structure somewhat in the outer disc, which is perhaps unexpected given the H$_2$CO column density map in \citetalias{Evans&Ilee2015}. Equally surprising is the relatively weak reflection of the underlying spiral structure in the CN integrated intensity map given that in \citetalias{Evans&Ilee2015} we concluded that CN was one of the strongest candidates for tracing non-axisymmetric structure in our disc model based on its abundance distribution. Moreover, we found the column density map of SO to be limited in extent in \citetalias{Evans&Ilee2015}, yet in the integrated flux maps here it traces as much of the disc as CO, albeit with slightly less contrast in the inner regions. Note, however, that sulfur-based chemistry is not currently fully understood, and so these results in particular should be taken with caution. 

\smallskip

We emphasise that \autoref{fig:integratedintensitymaps} suggests that a high contrast in abundance density does not necessarily translate into a high contrast in flux because of the complex nature of radiative transfer. As a result, it is crucial to perform radiative transfer calculations when one considers line flux imaging.

\subsubsection{Required sensitivities}
\label{sec:sensitivities}

\begin{figure*}
    \includegraphics[width=0.975\textwidth]{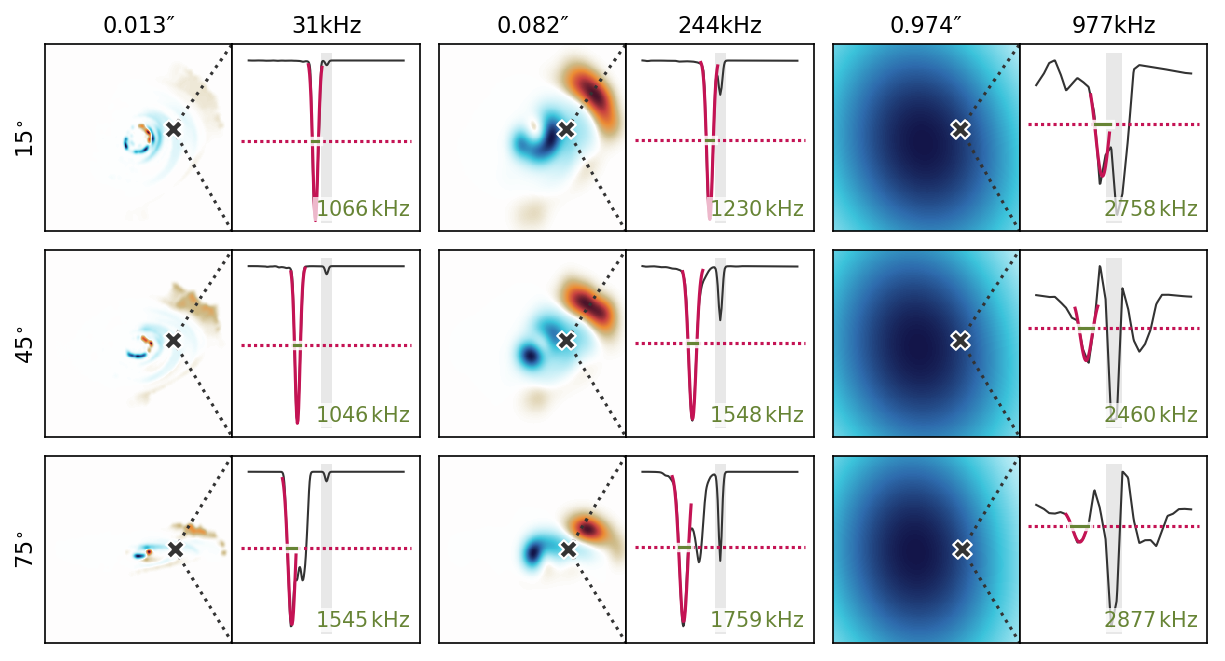}
    \caption{Depiction of the process undertaken to determine the required sensitivity for detection of line flux originating from spiral features in the disc model. Each panel shows, for a particular inclination, antenna configuration and channel width, the integrated intensity map with a pixel position within a spiral arm marked by a cross (left) and the line spectra extracted at this pixel position (right). In the spectra plots the solid purple line indicates a Gaussian fit, the dashed purple line indicates the half-maximum amplitude of this fit and the green line and accompanying text detail the FWHM, which is used to determine if the Gaussian fit can be resolved spectrally. The grey band indicates the envelope velocity range and if the Gaussian peak lies within this band then the line flux and envelope flux are indistinguishable. The line transition used here as an example is C$^{17}$O 3$\,\rightarrow\,$2.}
    \label{fig:sensitivityprocess}
\end{figure*}

From the perfect observations (i.e. no noise included) we extract the flux attributable to a typical spiral feature in our disc model, and determine the sensitivity required to detect this flux at the 4$\sigma$ level, which we take as the threshold for a reliable detection. To accomplish this we identify a pixel position within the spirals for a particular combination of molecular transition and disc inclination. We use the integrated intensity map with the smallest beamsize here because the spirals are the most distinguishable. We then extract the line spectra at this pixel and fit a Gaussian to the line feature originating from the spirals. We again start with the smallest beamsize, and also the smallest channel width in this case, so that we can accurately identify line features that originate within the spirals. From this Gaussian we determine the linewidth and half-maximum flux. The linewidth is compared to the channel width to determine if the line feature can be spectrally resolved, and we take the flux at half-maximum to ensure that enough of a peak can be detected and identified as line absorption (as is the case for spiral-originating flux). Using the same pixel position we then repeat this process as the beamsize and channel width are increased. We then change the disc inclination and repeat the pixel selection and subsequent process until all inclinations have been processed. Finally, we move on to a different molecular transition and repeat the entire sequence. \autoref{fig:sensitivityprocess} illustrates some steps of this process visually.

\smallskip

\begin{figure*}
    \includegraphics[width=0.975\textwidth]{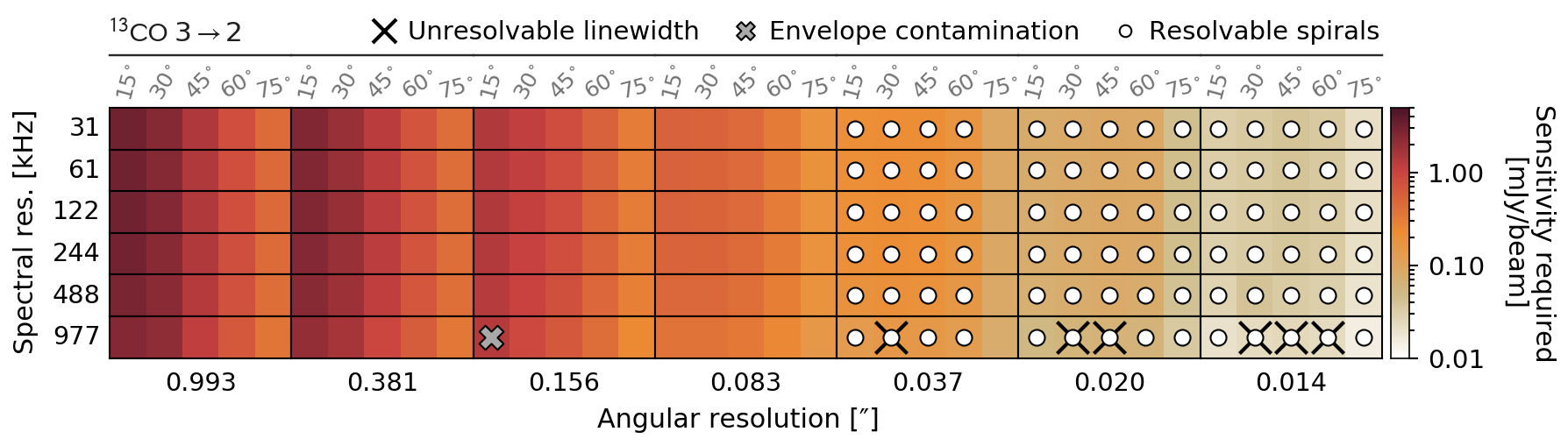}
    \caption{Sensitivity needed to detect $^{13}$CO 3$\,\rightarrow\,$2 line flux across different combinations of angular resolution (i.e. antenna configuration), spectral resolution and disc inclination (grey labels). The white circles indicate whether the spirals can be, in theory, spatially resolved, which is determined by eye from noise-free flux maps. The thick grey crosses indicate combinations that result in line spectra that are indistinguishable from the envelope. The thin black crosses indicate combinations where the Gaussian fitted to the line spectra (see \autoref{fig:sensitivityprocess}) is not spectrally resolvable.}
    \label{fig:13CO_3-2_SensitivityNeeded}
\end{figure*}

\begin{figure*}
    \includegraphics[width=0.975\textwidth]{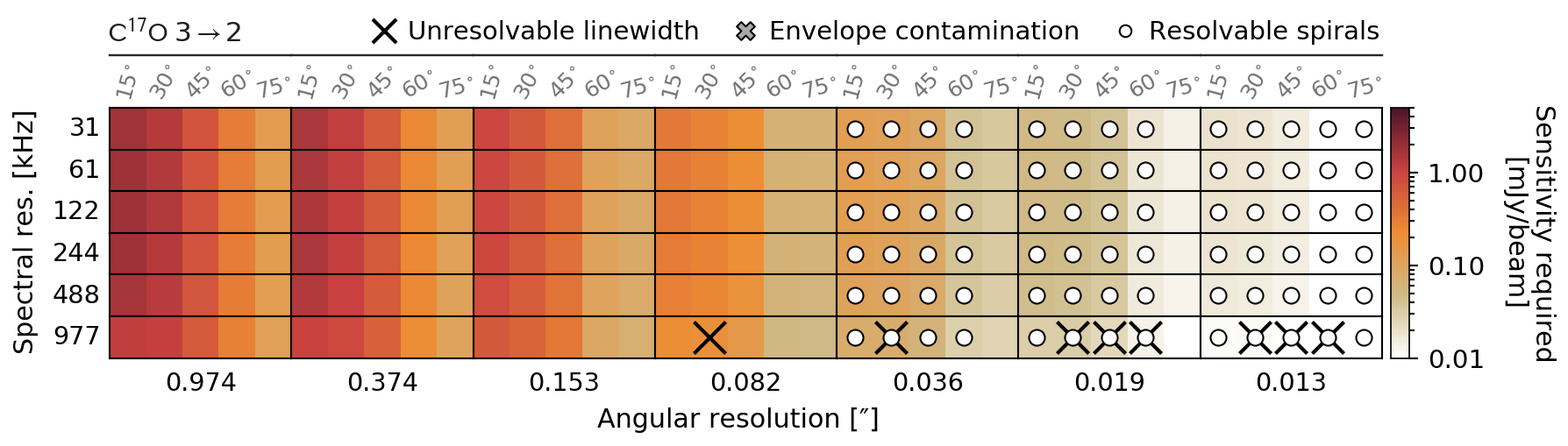}
    \caption{Same as \autoref{fig:13CO_3-2_SensitivityNeeded} but for C$^{17}$O 3$\,\rightarrow\,$2.}
    \label{fig:C17O_3-2_SensitivityNeeded}
\end{figure*}

\begin{figure*}
    \includegraphics[width=0.975\textwidth]{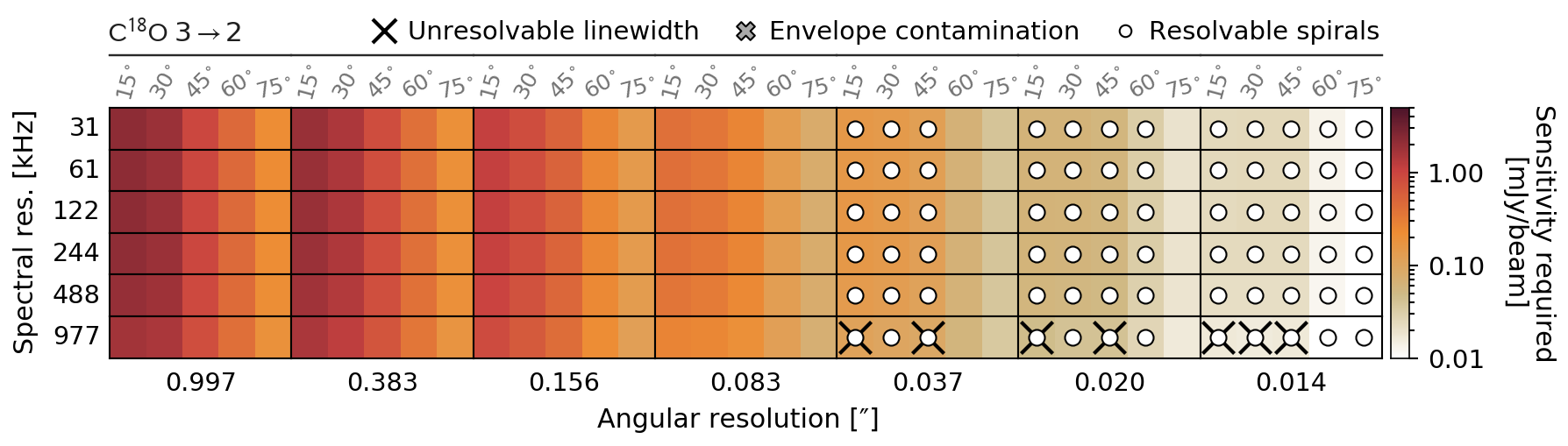}
    \caption{Same as \autoref{fig:13CO_3-2_SensitivityNeeded} but for C$^{18}$O 3$\,\rightarrow\,$2.}
    \label{fig:C18O_3-2_SensitivityNeeded}
\end{figure*}

\begin{figure*}
    \includegraphics[width=0.975\textwidth]{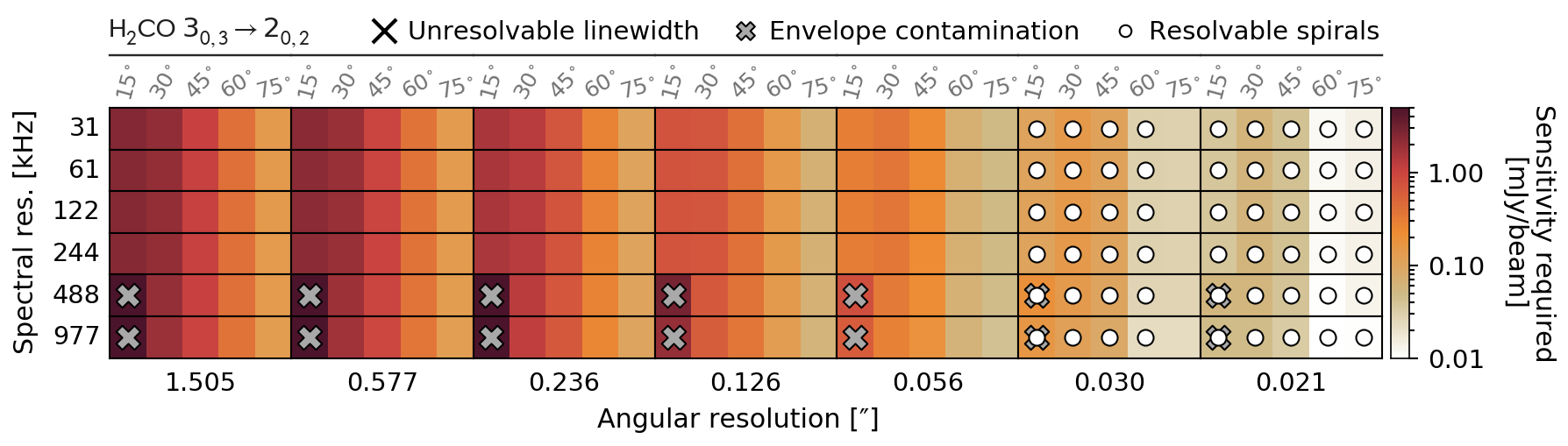}
    \caption{Same as \autoref{fig:13CO_3-2_SensitivityNeeded} but for H$_2$CO 3$_{0,3}\rightarrow2_{0,2}$.}
    \label{fig:H2CO_3-2_SensitivityNeeded}
\end{figure*}

\begin{figure*}
    \includegraphics[width=0.975\textwidth]{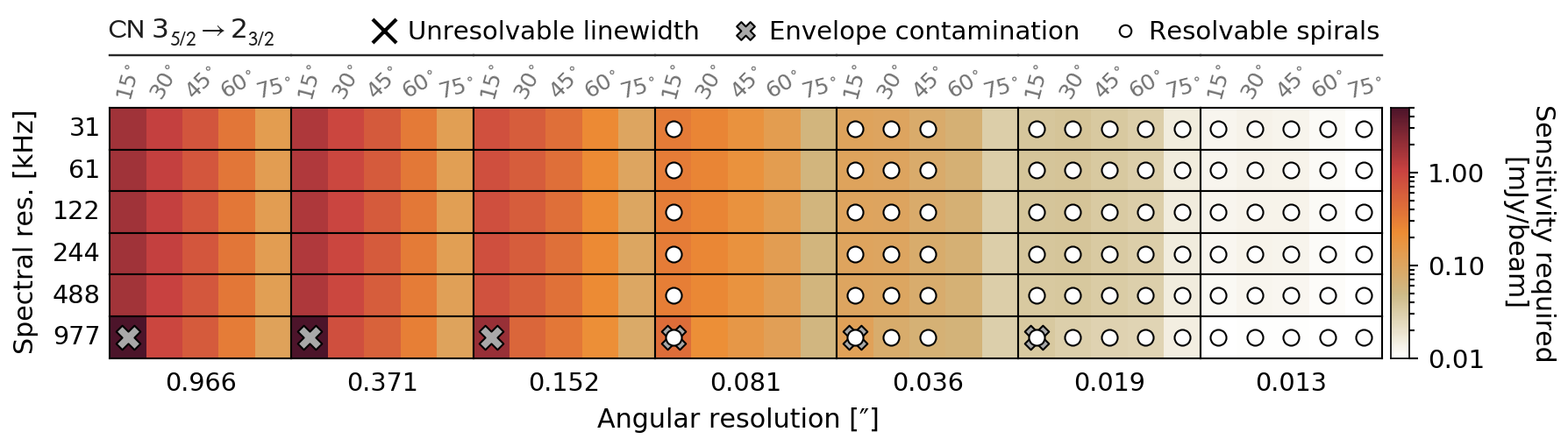}
    \caption{Same as \autoref{fig:13CO_3-2_SensitivityNeeded} but for CN 3$_{5/2}\rightarrow2_{3/2}$.}
    \label{fig:CN_3-2_SensitivityNeeded}
\end{figure*}

\begin{figure*}
    \includegraphics[width=0.975\textwidth]{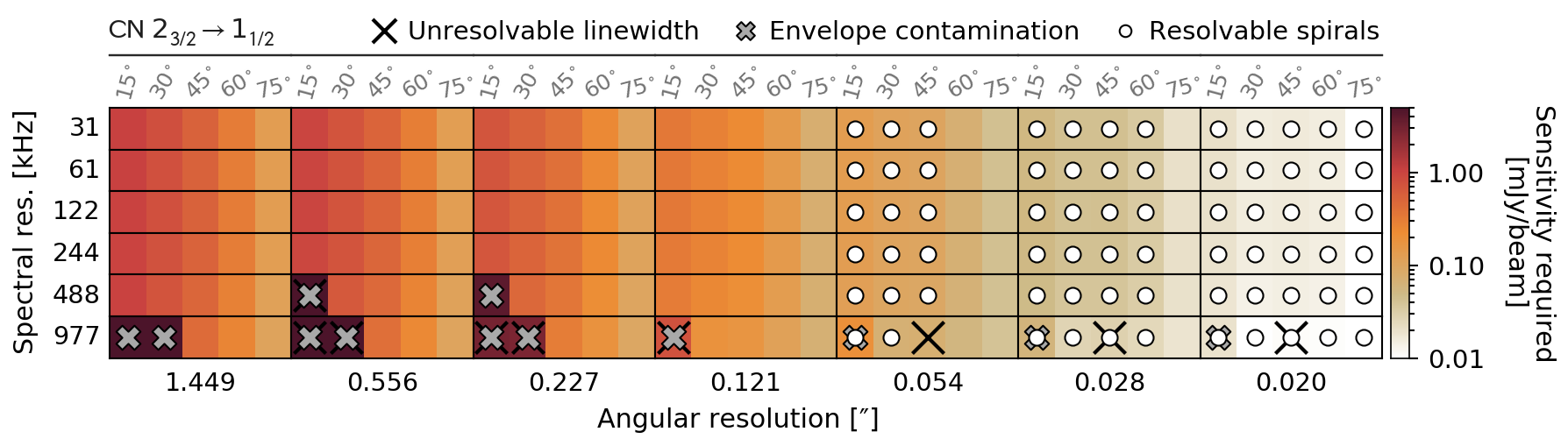}
    \caption{Same as \autoref{fig:13CO_3-2_SensitivityNeeded} but for CN 2$_{3/2}\rightarrow1_{1/2}$.}
    \label{fig:CN_2-1_SensitivityNeeded}
\end{figure*}

\begin{figure*}
    \includegraphics[width=0.975\textwidth]{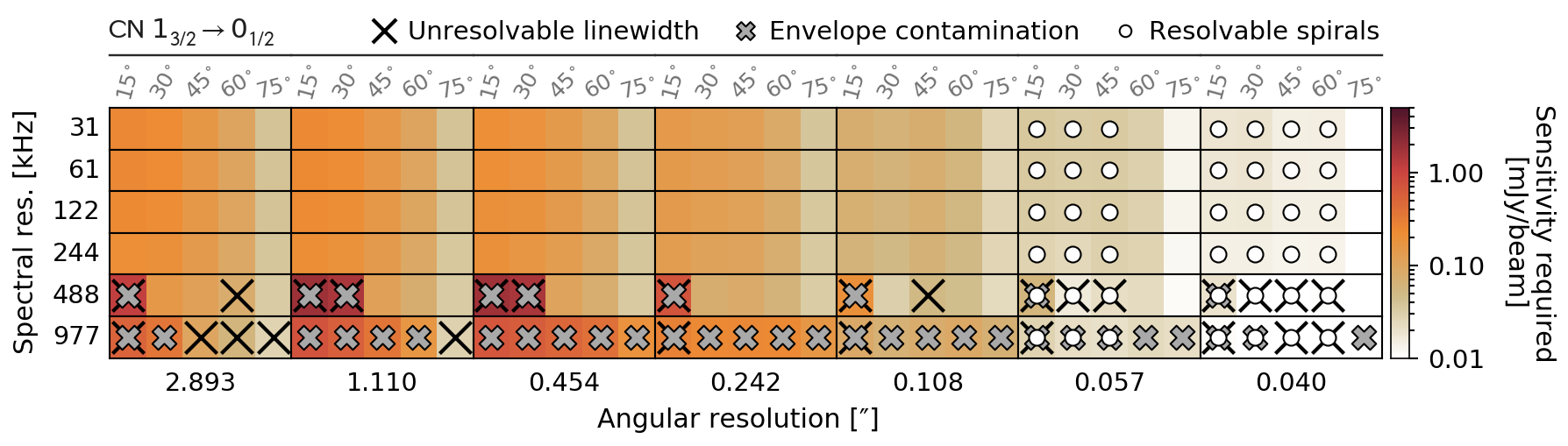}
    \caption{Same as \autoref{fig:13CO_3-2_SensitivityNeeded} but for CN 1$_{3/2}\rightarrow0_{1/2}$.}
    \label{fig:CN_1-0_SensitivityNeeded}
\end{figure*}

\begin{figure*}
    \includegraphics[width=0.975\textwidth]{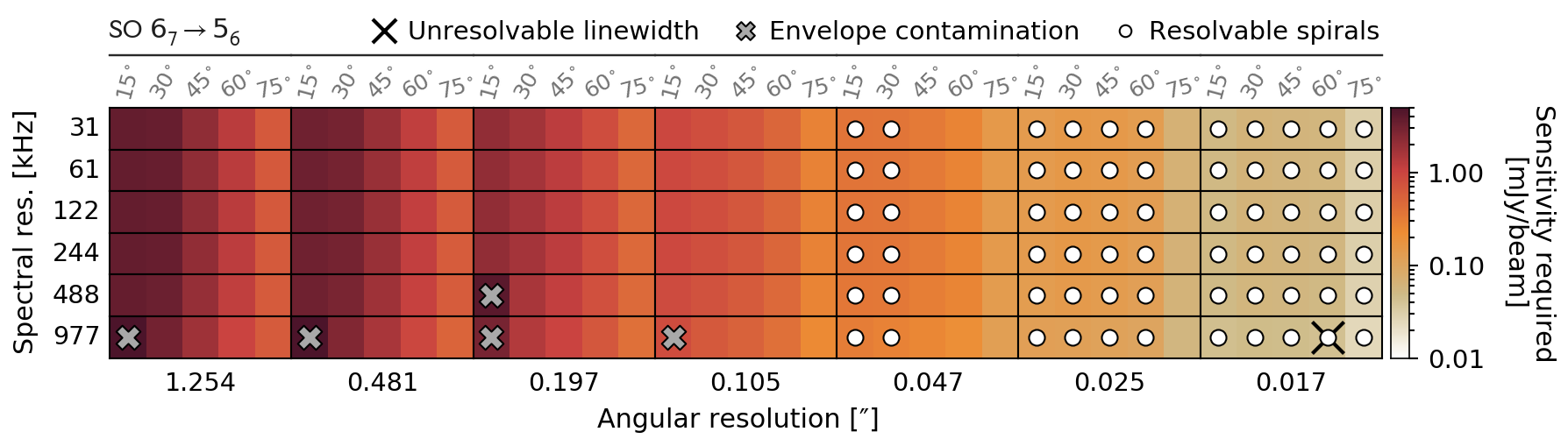}
    \caption{Same as \autoref{fig:13CO_3-2_SensitivityNeeded} but for SO $6_{7}\rightarrow5_6$.}
    \label{fig:SO_6-5_SensitivityNeeded}
\end{figure*}

We condense the extracted sensitivities into charts that depict the sensitivity required for each molecular transition we consider, which are shown in Figures \ref{fig:13CO_3-2_SensitivityNeeded}--\ref{fig:SO_6-5_SensitivityNeeded}. Within these charts is information pertaining to: whether the linewidth is unresolvable, i.e. whether the FWHM of the line is larger than the velocity resolution, indicated by a large black cross; whether the envelope substantially contaminates the line feature, indicated by a grey cross; and whether the spirals can be visually distinguished in an integrated intensity map, indicated by a white circle. Hence, wherever there is any type of cross in the charts, it means that detecting line flux from the disc spirals is unreliable for that particular combination of molecular transition, inclination, antenna configuration and channel width. Note that the particularly high sensitivities seen in some charts for the largest channel widths and beamsizes are because in those cases the envelope contribution can not be disentangled from the line flux (see the accompanying grey crosses), and so the Gaussian is fitted to a summation of line and envelope flux.


Figures \ref{fig:13CO_3-2_SensitivityNeeded}, \ref{fig:C17O_3-2_SensitivityNeeded} and \ref{fig:C18O_3-2_SensitivityNeeded} demonstrate that isotopologues of a species produce differing fluxes at similar frequencies. This is due to the difference in abundance of the isotopologues (because the only other factor that changes is the isotopic composition) where a larger number of absorbing molecules results in a larger optical depth and hence a deeper absorption feature, as shown by the purple line spectra in \autoref{fig:COComparisons}. However, the difference in peak flux recovered in the spiral regions is much less than the difference in abundance across the CO isotopologues; the change in abundance from $^{13}$CO to C$^{18}$O is $\approx$\,8 and from C$^{18}$O to C$^{17}$O is $\approx$\,4, whereas the change in peak flux is $\approx$\,2 and $\approx$\,1.5, respectively. The reason for this is that the dependence of line flux on the column density of absorbers, known as the `curve of growth', falls between $\sqrt{ln(N_i)}$ and $\sqrt{N_i}$ in the high optical depth regime. All of the CO isotopologues we consider occupy this regime in the disc spirals, and hence the difference in flux lies within this dependence range. Strictly speaking, the curve of growth defines how the equivalent width of a line depends on the column density, but because the line shape is dominated by a Gaussian, i.e. the Lorentz damping wings are unimportant, it is the peak of the absorption feature that changes. At very high optical depths, however, such as occurs in the innermost disc in CO, the line becomes saturated. As a result, the FWHM of the line feature changes, rather than the peak, in order to maintain the curve of growth dependence. This phenomenon is illustrated by the orange line spectra in \autoref{fig:COComparisons}.

\smallskip

Typically, for a particular temperature, the fractional population of levels decreases as the energy of the levels increases. This effect can broadly be seen across all molecules in \autoref{fig:levelpops}, albeit with some nuances due to the complex nature of some species' energy level structure. One of these nuances is the roughly constant fractional population of CN from $N$ = 0 to $N$ = 6, which we use to investigate the effect of transition frequency on flux detected, somewhat independently of the energy level populations. Figures \ref{fig:CN_3-2_SensitivityNeeded}--\ref{fig:CN_1-0_SensitivityNeeded} show that, in this case, the sensitivity required increases with transition frequency, which is expected because the optical depth increases with frequency, resulting in a deeper absorption feature at the same position within the disc model. This phenomenon is illustrated in \autoref{fig:CNComparisons}.

\subsubsection{Spatial resolution}
\label{sec:spatial}

The white circles in Figures \ref{fig:13CO_3-2_SensitivityNeeded}, \ref{fig:C17O_3-2_SensitivityNeeded} indicate whether or not the spiral features can be spatially resolved for each combination of observation parameters. However, it is important to note that the colour scale in the sensitivity charts indicates only whether a detection can be made; achieving the sensitivity level required to actually distinguish the spiral features visually is much more challenging, and likely not possible with ALMA across a large region of our parameter space.

\begin{figure}
    \includegraphics[width=0.475\textwidth]{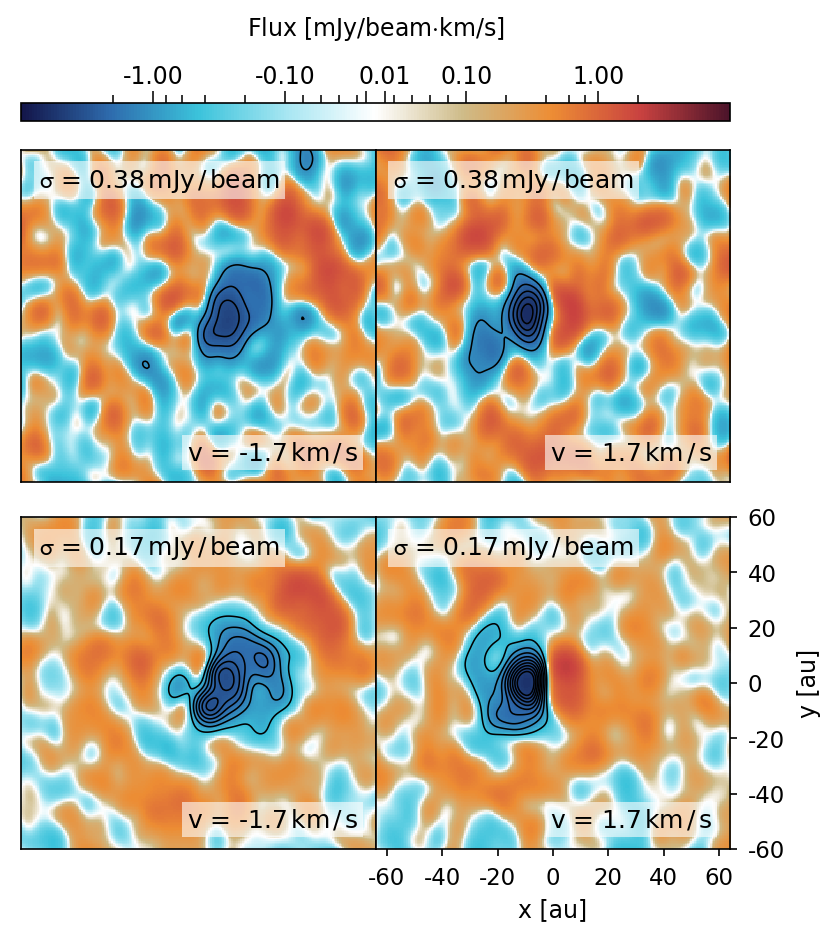}
    \caption{Channel maps for SO $6_{7}\rightarrow5_6$ indicating whether spiral structure can be visually distinguished for two different sensitivities; the lower sensitivity is the noise we determined is necessary to detect the spirals, whereas the higher sensitivity is the noise required to detect the contrast between and arm and inter-arm regions. The contours indicate levels of 4$\sigma$ and the bottom right panel indicates the physical scale of the observed disc model, which is constant across all transitions shown.}
    \label{fig:SO_6-5_SpatialResolving}
\end{figure}

For example, the lowest sensitivity required to detect spiral structure at a significant level (4$\sigma$) in our disc model is 0.38\,mJy\,beam$^{-1}$, which is recovered for SO 6$_{7}\rightarrow5_{6}$ (see \autoref{fig:SO_6-5_SensitivityNeeded}). As sensitivity decreases with spectral resolution, we use 977\,kHz and the ALMA Sensitivity Calculator with an optimal precipitable water vapour of 0.472\,mm to determine that an on-source observation time of 6.7\,hr is needed. The top panels of \autoref{fig:SO_6-5_SpatialResolving} show that, when we produce a synthetic observation under these conditions, no spiral structure is visually apparent. This is because the visual distinction of spiral features requires the ability to resolve the contrast between arm and inter-arm regions, and therefore the sensitivity needed is limited by the inter-arm flux. Adjusting the noise accordingly, spiral structure can indeed be seen at a 4$\sigma$ level in the resultant channel maps as expected, which is illustrated by \autoref{fig:SO_6-5_SpatialResolving}. The caveat here, however, is that now we require a sensitivity of 0.175\,mJy/beam, which implies an on-source time of 31 hours. Hence, spatially resolving spiral structure in a nearby, protosolar-like gravitationally unstable disc is potentially feasible, but will require a dedicated observation.

\subsubsection{Spectral resolution}
\label{sec:spectral}

If we relax the condition of spatially resolving the spiral structure, and instead focus on \textit{spectrally} resolving the spiral structure, then much shorter observation times are permissible. For example, the sensitivity required for a 4$\sigma$ detection of H$_2$CO 3$_{0,3}\rightarrow2_{0,2}$ in the spirals is approximately 2.3\,mJy\,beam$^{-1}$ when using a large beamsize (see \autoref{fig:H2CO_3-2_SensitivityNeeded}). As a narrow channel spacing is preferable for spectral resolution, we use a channel spacing of 31\,kHz with the ALMA Sensitivity Calculator, with automatic calculations of the precipitable water vapour and sky and system temperatures, and determine that this sensitivity requires an on-source observing time of $\approx$\,7.7\,hr. If we adopt a coarser velocity resolution of 244\,kHz, which is still sufficiently smaller than the linewidth across Figures \ref{fig:13CO_3-2_SensitivityNeeded}--\ref{fig:SO_6-5_SensitivityNeeded}, then an on-source observing time of $\approx$\,1.0\,hr is sufficient. This is firmly within the capabilities of ALMA and is consistent with the recent detection of molecules within low resolution observations of young protoplanetary discs \citep[e.g.][]{Tobin&Hartmann2012, Ohashi&Saigo2014, Sakai&Oya2014, Sakai&Oya2016, Tobin&Kratter2016}.

\smallskip

\begin{figure*}
    \includegraphics[width=0.975\textwidth]{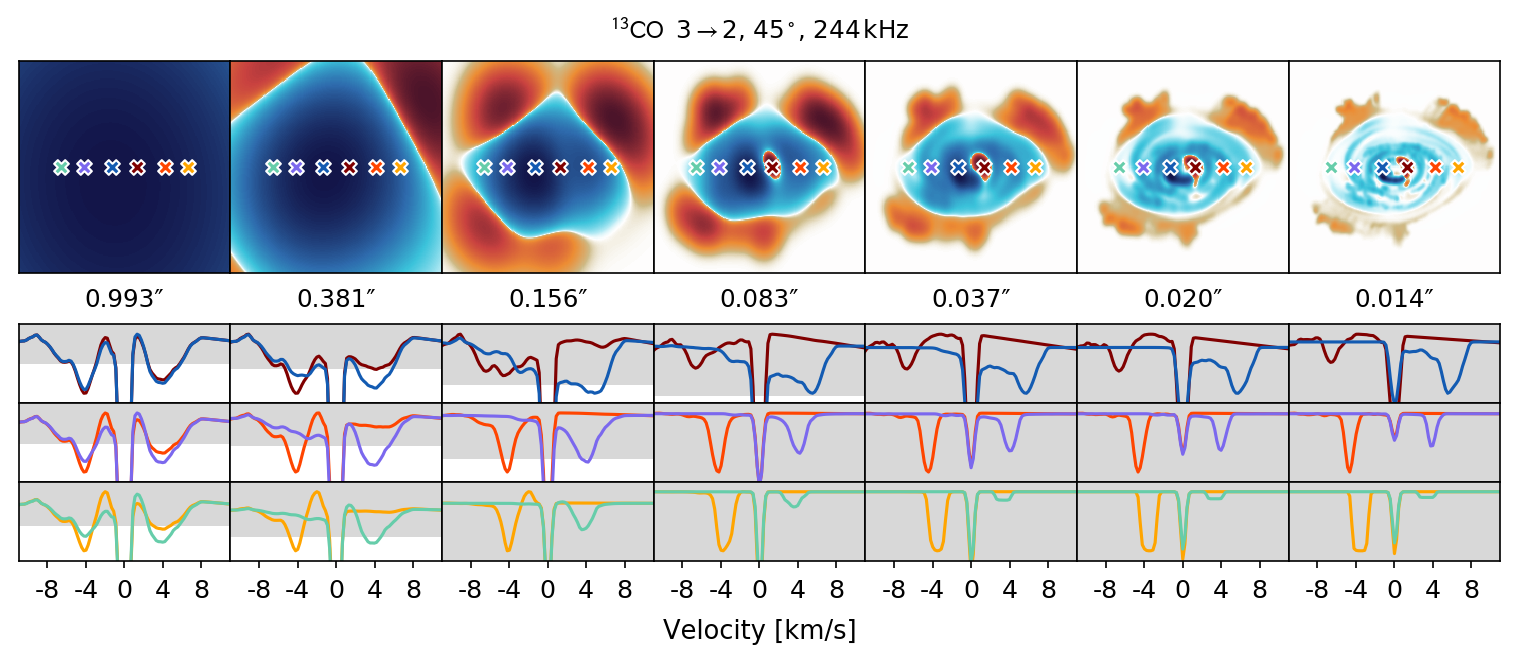}
    \caption{Integrated intensity maps of $^{13}$CO 3$\,\rightarrow\,$2 for different antenna configurations, with no noise included. Overlaid are crosses denoting the positions of molecular spectra extraction, and these extracted molecular spectra are displayed below the intensity maps for a spectral resolution of 244\,kHz, normalised against the non-envelope flux minima and maxima. The intention of the noise-free line spectra is to determine whether non-axisymmetric features persist across convolution with different beamsizes.}
    \label{fig:13CO_3-2_LineSpectra}
\end{figure*}

The pressing question here is whether the spiral structure can be distinguished spectrally from such observations. In order to investigate this we extract molecular line spectra for the species we consider, across the different antenna configurations, for a channel width of 244\,kHz. We extract the spectra from noise-free images at pixel positions in pairs that are symmetric about $x$ = 0, along the $y$ = 0 axis, in order to ensure a range of the disc morphology and structure is sampled. The intention is to assess whether line features can be distinguished with low angular resolution observations and, if so, whether these line features can be attributed to the spiral structure (as is the case for high angular resolution observations) or instead are tainted by convolution. Finally, we compare spectra taken from symmetrically positioned pixels in order to see if non-axisymmetric structure can be distinguished. \autoref{fig:13CO_3-2_LineSpectra} shows the integrated intensity maps, where the integration is performed from -12 to -1\,km\,s$^{-1}$ and 1 to 12\,km\,s$^{-1}$ in order to avoid the envelope contribution, with the pixel positions overplotted for $^{13}$CO, at an inclination of 45$^\circ$, along with line spectra plots. This is intended to serve as a representative example but the plots for all of the transitions can be found at \url{https://goo.gl/AVkHRj}.

As can be seen in \autoref{fig:13CO_3-2_LineSpectra}, the line spectra convolved with large beamsizes are not necessarily faithful to the underlying spectral structure. This is because at the highest angular resolution, spectra taken from the left and right sides of the disc show a line feature either at positive or negative velocities (excluding the envelope centered at 0\,km\,s$^{-1}$) due to either being red or blue-shifted in respect to the observer. At the lowest angular resolutions, however, all spectra exhibit features at both positive and negative velocities. This finding applies to all of the transitions we consider, which means that in order to achieve a compromise between flux and spectral accuracy, intermediate beamsizes (in terms of the beamsizes available with ALMA) are required.

\smallskip

At intermediate beamsizes, \autoref{fig:13CO_3-2_LineSpectra} shows that overall the relations between line peaks originating in the blue-shifted and red-shifted portions of the disc are maintained from the highest angular resolution observations. Again this is true for all of the molecular line transitions we consider, and so intermediate angular resolution spectra could be used, in theory, to infer the presence of non-axisymmetric structure in a young, embedded disc. This is, however, dependent on the noise level that can be reached in an observation.

\begin{figure}
    \includegraphics[width=0.475\textwidth]{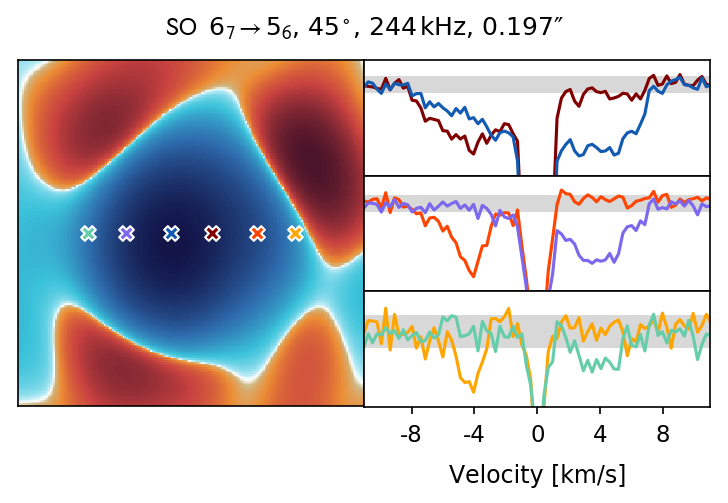}
    \caption{Integrated intensity map of SO $6_{7}\rightarrow5_6$ with an on-source observation time of 2.5\,hr. Overlaid are crosses denoting the positions of molecular spectra extraction, and these extracted molecular spectra are displayed below the intensity map for a spectral resolution of 244\,kHz, normalised against the non-envelope flux minima and maxima. The grey regions in the spectra plots denote the noise level, which is 1.24\,mJy/beam.}
    \label{fig:SO_6-5_out13_LineSpectra}
\end{figure}

In order to demonstrate that this hypothesis does hold under realistic noise levels, we again use the ALMA Sensitivity Calculator, with optimal atmospheric conditions, in order to relate the sensitivity required for spiral detection (see Figures \ref{fig:13CO_3-2_SensitivityNeeded}--\ref{fig:SO_6-5_SensitivityNeeded}) to a required on-source observation time. For SO $6_{7}\rightarrow5_6$, for example, the sensitivity required to detect spiral features in our disc model is 1.24\,mJy/beam when using an intermediate beamsize of 0.197\,arcsec and a spectral resolution of 244\,kHz. \autoref{fig:SO_6-5_out13_LineSpectra} shows the line spectra extracted from a realistic synthetic observation performed under these conditions, and as can be seen, the non-axisymmetric features are recovered as expected. As a result, we find that relatively short-duration ALMA observations should be perfectly capable of extracting non-axisymmetric features in young, protosolar-like embedded discs. Note, though, that this outcome becomes less reliable as the disc approaches an edge-on inclination due to an increased spectra complexity caused by the increased amount of material along the line-of-sight.

\smallskip

We have implemented a simplistic envelope model when performing the radiative transfer calculations in this paper, which has permitted us to straightforwardly isolate the envelope contribution in most line spectra. However, several molecular species have been observed to trace material flowing out from discs and accreting onto discs, which could contaminate the recovered line spectra in more complex manners. A method of combatting this is to use a high spectral resolution so that the different kinematic components can be disentangled, but, as the on-source observation time increases with spectral resolution, a compromise will most likely have to be made. An alternative method is to focus on species that do not typically trace envelope accretion or outflows, such as C$^{17}$O and OCS.

\subsubsection{Position-velocity diagrams}
\label{sec:pvdiagramssec}

Whilst we have shown that non-axisymmetric structure could be inferred from relatively low angular resolution observations via spectral analysis, concluding that this structure is spiral in nature, and therefore possibly driven by GIs, requires more evidence. A particularly strong indicator of spiral structure in Galactic simulations is the detection of finger-like features in a position-velocity (PV) diagram \citep[see e.g.][]{Bissantz&Englmaier2003, Rodriguez&Combes2008, Li&Gerhard2016}, and this can also be applied to GI-active protoplanetary discs \citep{Douglas&Caselli2013}. Therefore, we produce noiseless PV diagrams for all of the molecular transitions we consider in this paper, across our entire parameter space, in order to determine if identifying spiral structure is theoretically possible. These can be found at \url{https://goo.gl/AVkHRj}, but here we show the PV diagram for SO 6$_{7}\rightarrow5_{6}$ across a subset of angular resolutions as a representative example.

\begin{figure*}
    \centering
    \includegraphics[width=0.975\textwidth]{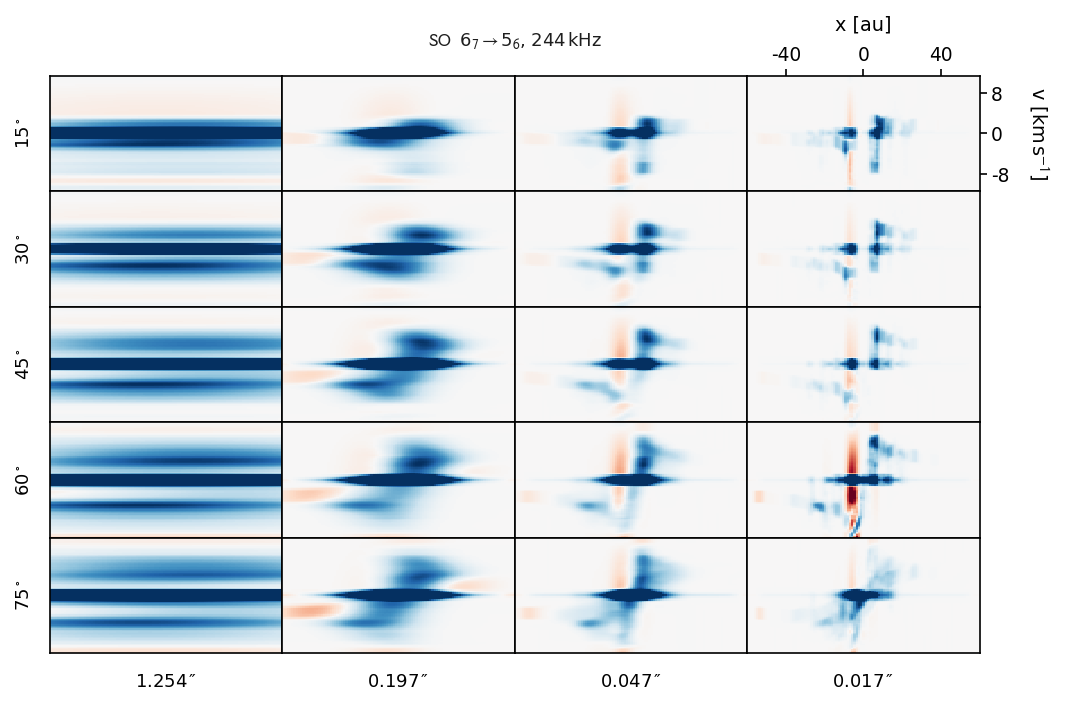}
    \caption{Position-velocity diagram for SO 6$_{7}\rightarrow5_{6}$ across different disc inclinations and beamsizes, for a spectral resolution of 244\,kHz. The top right panel indicates the position and velocity ranges of the observed disc model, which are constant across all panels.}
    \label{fig:SO_6-5_PVDiagram}
\end{figure*}

\autoref{fig:SO_6-5_PVDiagram} shows the position-velocity diagram for SO 6$_{7}\rightarrow5_{6}$ across various inclinations and antenna configurations, for a channel width of 244\,kHz. As can be seen, near edge-on (90$^\circ$) inclinations are required to distinguish finger-like structure in the PV diagram. Moreover, in general high angular resolutions are also required, which, when cross-referencing with the sensitivity chart (\autoref{fig:SO_6-5_SensitivityNeeded}) suggests that if the spirals cannot be spatially resolved, then they cannot be identified in PV diagrams either. However, a caveat to this exists at intermediate beamsizes (i.e. 0.197--0.105\,arcsec for SO 6$_{7}\rightarrow5_{6}$) because the high-contrast individual fingers seen at the highest angular resolutions, which directly trace the spiral structure, are essentially blurred together at lower angular resolution. This results in asymmetric absorption features extending to high positive and negative velocities from the center of the PV diagram, which suggests that non-axisymmetric structure exists within the disc. This means that PV diagrams of intermediate-beamsize ALMA observations could be used to infer the presence of GI-driven spiral structure, even if the spirals cannot be resolved spatially.

\smallskip

As with the Sections \ref{sec:spatial} and \ref{sec:spectral}, this result is intended to foremost serve as a theoretical exploration into the feasibility of detecting spiral structure. As discussed, in real observations, it is only possible if the flux signal exceeds the sensitivity, which is dependent on the on-source integration time. The sensitivity charts (see Section \ref{sec:sensitivities}) can be used with the ALMA Sensitivity Calculator to approximate the on-source time required, which then indicates whether an observation would be possible. In terms of PV diagrams, for SO 6$_{7}\rightarrow5_{6}$, for example, inferring the presence of spiral structure from the bottom right panel of \autoref{fig:SO_6-5_PVDiagram} requires a sensitivity of 0.02\,mJy/beam. Under optimal atmospheric conditions, this sensitivity requires an on-source observation time of just over a year, which is clearly not feasible. If instead we use a larger beamsize, such as 0.197\,arcsec, then spiral structure can still be inferred and the on-source time is reduced to 19\,hr, which could potentially be achieved given a dedicated observation. This means that confirming spiral structure from PV diagrams of young, embedded protosolar-like discs is currently extremely challenging, but potentially affords shorter observations when compared to spatially resolving spiral features (with the caveat that the disc must be highly inclined).

\subsection{Significance of absorption features}

All spectroscopic features detected in non-edge-on discs observed around T Tauri stars are emission features \citep[e.g.][]{Williams&Cieza2011, Ansdell&Williams2017}. This is because an irradiation-dominated disc has a surface temperature that is higher than its interior temperature, even when viscous heating is included in models via the $\alpha$-disc prescription. Here, however, we use a disc model that is viscosity-dominated because we are focusing on the earliest and most observationally-challenging stages of disc formation. As a result, we recover cooler surface layers obscuring a hot midplane, rather than the warm surface layer seen in observations and simulations \citep[see e.g.][and references within]{Henning&Semenov2013}, which means the majority of our disc model is seen in absorption. From this we propose that absorption could be a signature of a young, embedded protoplanetary disc dominated by viscous heating. 

\smallskip

Because the temperature falls off as roughly $r^{-0.5}$ in the $\alpha$-disc prescription, the detection of absorption across the majority of the extent of the disc would imply that this viscous heating is global in nature. As we show in \citetalias{Evans&Ilee2015}, gravitational instability can drive global shock heating throughout the disc that warms the midplane to large radii. Therefore, although confirming non-axisymmetric structure is certainly a challenge in young, embedded discs, inferring the development of gravitational instability from low resolution absorption signatures may still be possible.

\smallskip

However, an unwanted consequence of absorption features is that, at low angular resolution, emitting and absorbing regions can be blended together, which will obfuscate the observed flux. Consequently, these opposing regions could destructively interfere, greatly reducing the flux that is received. This could then indicate that the observed system possesses a much weaker flux than it really does, which could essentially mimic the effects of depletion. As the study of depletion of volatile species is an important topic in protoplanetary discs \citep[e.g.][]{Kama&Bruderer2016a, Kama&Bruderer2016b}, ensuring reliable fluxes are recovered from observations is of pivotal importance. Thus, care should be taken to account for the effect of overlapping absorbing and emitting regions when analysing integrated line intensity maps and line spectra.

\begin{figure*}
    \centering
    \includegraphics[width=0.85\textwidth]{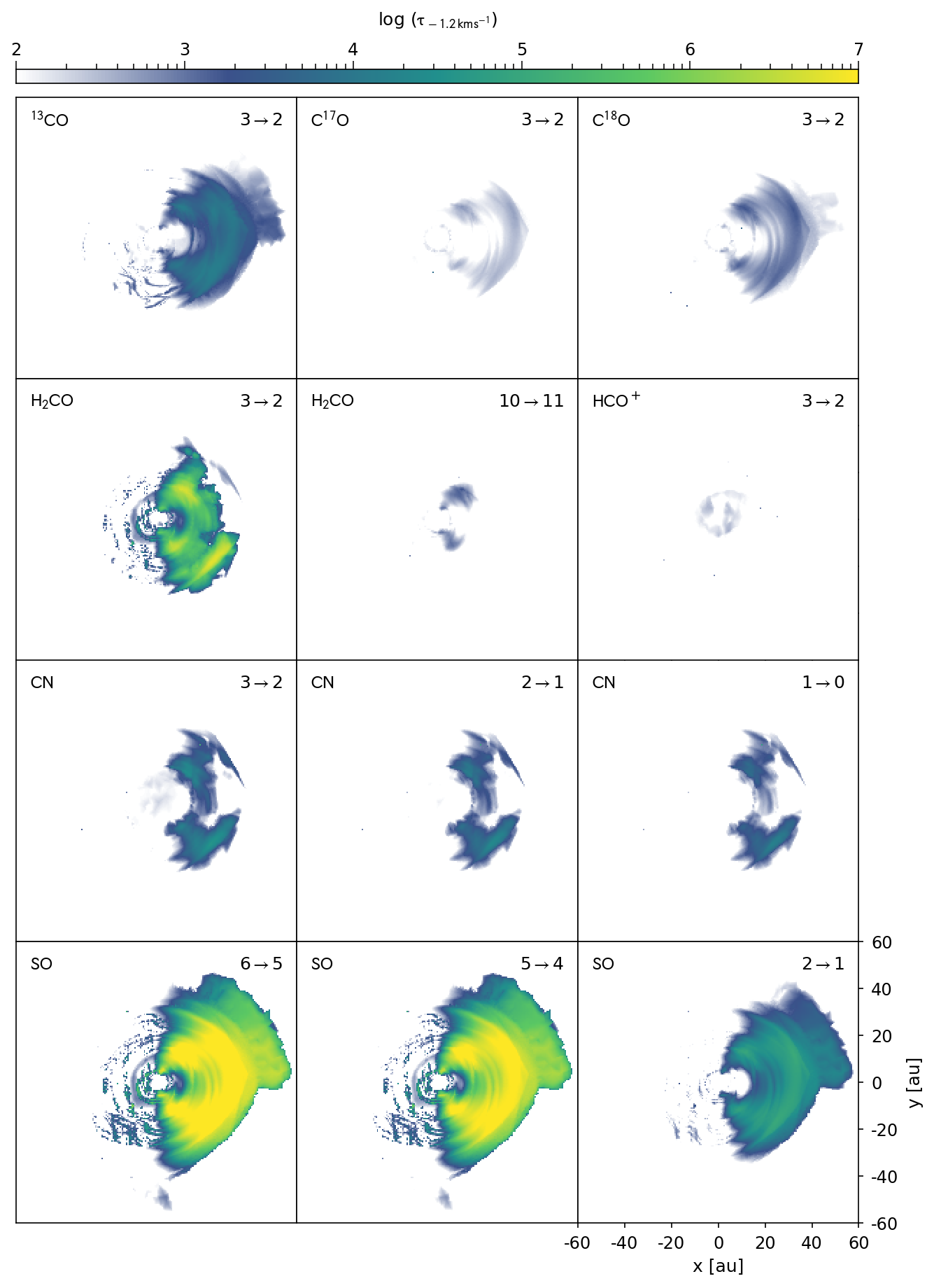}
    \caption{Optical depth at $v$ = -1.2\,km\,s$^{-1}$ for each molecular transition, at an inclination of 15$^\circ$, calculated from the \lime\ images directly. The bottom right panel indicates the physical scale of the disc model, which is constant across all transitions shown.}
    \label{fig:taumaps}
\end{figure*}

The absorption features we recover from our disc model originate in the surface layers of the disc. This is due to, firstly, the high optical depths of the continuum even at the lowest frequencies we consider (see \citetalias{Evans&Ilee2017}), and secondly, because the optical depths of all the transitions we feature, including less abundant CO isotopologues, are very large as shown in \autoref{fig:taumaps}. Consequently, it is important to emphasise that such high optical depths will lead to systematic underestimates of column densities and disc masses. Recently, \citet{Yu&Evans2017} found high optical depths in a (gravitationally stable) 0.03\,M$_\odot$ protoplanetary disc model that lead to underestimated disc masses even in the most optically thin CO isotopologue they consider, C$^{17}$O. Moreover, \citet{Miotello&vanDishoeck2017} found that the optical depth effects will be larger (and more complex) for more massive discs, and hence the mass will be more severely underestimated.

\smallskip

We note that in this work we have assumed that the gas and dust temperatures are equivalent within the model disc. However, if this assumption is inaccurate and the dust temperature is significantly lower than the gas temperature, then molecular line fluxes may be seen in emission, rather than absorption as we have shown. However, this quandary essentially simplifies to whether the kinetic temperature is equivalent to the excitation temperature, which is the condition for LTE. As we have shown in Section \ref{sec:nonlte} that LTE is appropriate for our disc model across the low energy transitions, our assumption of thermal balance should be valid \citep[see also][Appendix A]{Yu&Evans2017}. If higher energy transitions are observed, or the disc is assumed to be less embedded, then the gas and dust temperatures may become decoupled, resulting in the disc being seen in emission.

\subsection{Comparison with other studies}

\citet{Douglas&Caselli2013} explored the observability of spiral features in a more massive, gravitationally unstable, disc (0.39M$_\odot$; \citealt{Ilee&Caselli2011}) than we consider. The authors conclude that spiral features are detectable in absorption across a large parameter space. However, the sensitivities the authors adopt were calculated for a bandwidth of 1.875\,GHz, which is the available bandwidth of the basebands within the ALMA band receivers, rather than the expected linewidth of each transition as is appropriate (approximately 350--450\,kHz across the frequencies listed in their Table 1). Hence, their sensitivities are significantly higher than we find, meaning \citet{Douglas&Caselli2013} report much shorter observation times are required than we report in this paper. Nevertheless, our results agree with the detection of a disc in absorption and confirm that, in principle, spiral features can be inferred from observations of protoplanetary discs both spatially and spectrally if the signal-to-noise ratio is high enough.

\smallskip

\citet{Seifried&Sanchez-Monge2016} used 3D magnetohydrodynamics to simulate the collapsing of two different sized molecular cores into protoplanetary discs systems. Their model 2 features a Bonnor-Ebert sphere envelope, as we have used, and evolves to form a single 0.62\,M$_\odot$ protostar and a disc with an approximately 70\,au radius, both of which are comparable to our disc model. The authors use \radmc\ to produce line images for different molecular transitions, and then use \casa\ to synthesise observations of these line images across different observing times, angular resolutions, inclinations and weather conditions. Overall, the authors conclude that Class 0 discs should be observable in numerous molecular transitions, across a wide range of observational parameters. However, no structural detail is apparent in the integrated intensity maps for their model 2, although this is partly due to the fact that the model 2 disc remains gravitationally stable, and hence does not develop strong spiral features. On the other hand, the model 1 disc, which is most likely much more massive than our disc model, does contain non-axisymmetric structure, but this structure is only partially reflected in the integrated intensity maps. Therefore, this supports our findings that spatially resolving spiral features is extremely challenging in protosolar-like discs.

\smallskip

The integrated intensity maps in \citet{Seifried&Sanchez-Monge2016} all show line flux in emission, whereas for our disc model we see the majority of line flux in absorption. A potential explanation for this discrepancy is that the authors assume constant abundances for all species they consider, whereas we have shown in \citetalias{Evans&Ilee2015} that spiral shocks have a significant effect on the abundances of many molecular species. However, this should only affect the contrast between the spiral and inter-spiral regions in their intensity maps, rather than convert emission to absorption. Moreover, the authors adopt lower abundances, which are more in-keeping with our peak abundances, and find similar results in emission. If the \citet{Seifried&Sanchez-Monge2016} disc models were not GI-active, then this would offer a relatively simple alternate explanation. However, the authors report that their disc models are gravitationally unstable and do indeed develop the signature spiral structure. Therefore, we instead posit that the amplitudes of the spiral density waves in their disc models are lower than in the disc model we consider, which means the midplane regions are likely not heated as significantly by global shocks; hence the upper layers are warmer and the discs are seen in emission. These lower amplitudes may result from suppression of the GI activity caused by the inclusion of a magnetic field in the \citet{Seifried&Sanchez-Monge2016} disc models. But, as the interaction between MRI turbulence and GI turbulence is currently largely unknown, more simulations are required to constrain the conditions under which young, embedded protoplanetary discs will be seen in absorption.

\section{Conclusions}
\label{sec:conclusions}

\subsection{Radiative transfer calculations}

We have produced line flux maps of a protoplanetary disc model using \lime\ and in Section \ref{sec:limeimages} have explored how to attain accurate images. This involved making adjustments to the default sampling behaviour of \lime, comparing LTE treatment to non-LTE treatment, and determining the frequencies that reliably reflect the disc composition. The main results from this section are:

\begin{itemize}[leftmargin=0.0cm, itemindent=1.0cm]

\item Specialised grid construction can improve the accuracy of molecular line images produced with \lime. We find that a custom sampling routine affords the most convergent line flux images for our gravitationally unstable disc model, which combines rectilinear sampling of the disc with the restriction of grid points at $\tau$ \textgreater\ 3.

\item A convergence test should be used to find the number of grid points necessary to produce an accurate line flux image. This is model dependent but for our young, gravitationally unstable disc model, we find that a minimum of 1 $\times$ 10$^6$ points should be used. Note, however, that more grid points are always preferable if the computational cost can be afforded. Incidentally, this number of points is larger than is necessary in the continuum due to the increased complexity of radiative transfer calculations when considering line flux.

\item Performing radiative transfer calculations for line fluxes can be extremely computationally intensive due to the coupled nature of the intensity, absorption and emission coefficients, and the level populations. A significant speed-up can be made if LTE is adopted, but the validity of this approximation should be checked by comparing the number density of the emitting regions to the critical density (or effective excitation density). One can also compare the LTE run to a full non-LTE run in order to ensure the results are not significantly divergent; we find that non-LTE and LTE runs are congruent, and hence LTE is an accurate approximation for our disc model.

\item The frequency of a spectral feature can significantly affect the location of the observable emitting regions. At large frequencies the optical depth is higher, which means that less of the disc material is probed by observations. Consequently, the observed flux will originate from nearer the surface of the disc and hence closer to the critical density boundary, which degrades the applicability of LTE. Therefore we recommend observing embedded protoplanetary discs at low frequencies.

\item The resolution and extent of the abundances should be sufficient to encompass the emitting regions, as otherwise the reliability of line images can be questioned. This could have implications for simulations that distribute fluid elements by mass or density because the emitting regions can be located in low density regions, particularly at large frequencies, and hence may not be well sampled.

\end{itemize}

\subsection{Synthetic observations}

In Section \ref{sec:obs} we produced noiseless synthetic observations of our disc model by convolving the \lime\ images with a 2D Gaussian beam, across a large parameter space consisting of different molecular transitions, antenna configurations, spectral resolutions and disc inclinations. We assumed our observations are perfect, i.e. no noise present, in order to determine the sensitivity that is required for a detection. Our model represents a 0.17M$_\odot$ protoplanetary disc that surrounds a 0.8M$_\odot$ protostar likely to evolve into a Solar-like star. As a result our work may be indicative of observations of an object similar to our early Solar System. The main results from this section are as follows:

\begin{itemize}[leftmargin=0.0cm, itemindent=1.0cm]

\item All transitions we consider trace spirals to an extent in noiseless observations with high angular resolutions. However, reflection of spiral structure in integrated line intensity maps does not necessarily correlate with the underlying distribution of the species. Therefore, it is imperative that radiative transfer calculations are performed when considering line emission and absorption.

\item The gravitational instability within our model disc drives spiral density waves that shock-heat the disc material globally. As a result, because we assume the disc is heavily embedded, we find the disc is comprised of cooler surface layers obscuring a hot midplane. Consequently, the disc is seen primarily in absorption across all transitions we consider. Therefore, absorption, which can be attributed to infalling material, could also be a signature of GI in Class 0 sources.

\item The fact that our gravitationally unstable disc is seen in absorption could introduce complications for low angular resolution observations. This is because emitting and absorbing regions for a particular transition could blend together, reducing the overall flux that is detected. This could then affect further analysis and lead to erroneous conclusions such as the presence of significant depletion in molecular species.

\item All the molecular transitions we consider are substantially optically thick ($\tau$ \textgreater\ 10$^2$) in our disc model, even at the lowest frequency implemented (113\,GHz; Band 3 of ALMA) and for the least abundant molecule (HCO$^+$). Therefore, the flux detected is only tracing the upper layers of the disc, which means that estimates of column densities and disc masses derived from line emission or absorption in young, embedded sources will most likely be significantly underestimated.

\item The sensitivity required to detect absorption originating in spirals increases with frequency due to the increase in optical depth. The sensitivity also increases with abundance, albeit in a non-linear way determined by the `curve of growth'. Therefore, higher frequency transitions of more abundant isotopologues should afford easier absorption flux detections in young, embedded gravitationally unstable discs. However, the increased optical depths mean that less of the disc material is probed, which will lead to even larger systematic underestimates of column densities and therefore disc masses.

\item The sensitivity required for a spiral feature detection does not depend on the spectral resolution used, assuming the spectral resolution is smaller than the linewidth. The caveat to this is if the observed line spectrum is contaminated by the envelope flux contribution, which we found occurred for most species when using large angular resolutions in conjunction with large velocity resolutions. Therefore, because on-source observing time increases as channel spacing decreases, we recommend to observe young protoplanetary discs with a spectral resolution of 244--488\,kHz.

\item Spatially resolving the spiral structure in a GI-driven, protosolar-like disc is possible using molecular lines when the beamsize is comparable to the width of the spiral arms, i.e. at high angular resolutions. However, in this case, the sensitivities required for a significant detection result in a minimum on-source time of 30\,hr. Therefore, this technique may only be appropriate if ALMA changes to a deep-scan mode in the future.

\item Spectrally resolving the spiral structure in a GI-driven, protosolar-like disc may be possible if intermediate beamsizes (e.g. 0.15\,arcsec for $^{13}$CO 3$\,\rightarrow\,$2) are adopted; at lower angular resolutions the line spectra are altered too much by convolution. Non-axisymmetric structure can be inferred from detected spectra if symmetric positions within the disc show different peak amplitudes, which is possible given a relatively short minimum on-source observation time of 1\,hr. Therefore, ALMA should be readily able to spectrally detect non-axisymmetric structure in young, protosolar-like discs. Note, however, that it is not possible from these results to reliably conclude that the structure is GI-driven.

\item Direct confirmation of GI-driven spiral structure is achievable via PV diagrams with an source-time of 19\,hr if intermediate beamsizes are adopted. This is because, for nearly edge-on inclinations, finger-like features can be identified that directly correlate to spiral structures. Hence, with current instruments, PV diagrams probably offer the most promising means for confirming the presence of GI-driven spiral structure in young, embedded protoplanetary discs.

\end{itemize}

In this series of papers we have investigated the evolution of a gravitationally unstable disc within a Class 0 object that may be analogous to our early Solar System. We have performed 3D modelling of the dynamics and chemistry, followed by synthesis of accurate continuum and line observations, which have allowed us to study the connection between hydrodynamic instabilities, subsequent chemical evolution, and the resulting changes in observational characteristics for this protosolar-like system. We have shown that GI-driven shocks have a transient and permanent effect on disc structure and disc chemistry, and produce a chemically rich midplane in stark contrast to the `gas-phase desert' recovered in simulations of more evolved discs \citepalias[see][]{Evans&Ilee2015}. 

\smallskip

The spiral features in our gravitationally unstable disc model can be readily detected by ALMA in the continuum \citepalias[see][]{Evans&Ilee2017}, although these dust features are likely indistinguishable from alternate mechanisms that produce spiral structure, such as unseen companions \citep[see e.g.][]{Meru&Juhasz2017}; hence the conclusion of GI as the driving mechanism is dubious. However, if continuum observations can be complemented with line images that reveal coincident non-axisymmetric structure, which we have shown in this paper is possible with ALMA observations of low frequency line transitions, then this implies that both the dust temperatures and molecular abundances are locally enhanced. In this case, a much stronger argument can be made for the existence of global-scale spiral shocks that drive turbulence in a disc, likely caused by GI.

\smallskip

In the near future, the continual advancement in angular resolution and sensitivity of submillimetre interferometers such as ALMA, and the introduction of new near/mid-infrared instrumentation such as MIRI (Mid-InfraRed Instrument) on JWST (James Webb Space Telescope, to be launched in 2019) and the proposed European Extremely Large Telescope (E-ELT, first light planned for 2024), and new radio instrumentation like the proposed SKA (Square Kilometer Array, phase one operations scheduled for 2020) will increase the likelihood of unequivocally confirming the presence of gravitational instabilities in young, embedded protoplanetary discs. And, whilst this goal is challenging to accomplish, it is certainly worthwhile as it will enhance our understanding of the early stages of protoplanetary disc evolution, which will consequently shape the knowledge of later processes such as planet formation and the production of pre-biotic molecules that are direct descendants of life. 

\section{Acknowledgements}

We would like to thank the anonymous referee for their constructive comments that have significantly improved the clarity of this manuscript. MGE gratefully acknowledges a studentship from the European Research Council (ERC; project PALs 320620). TWH and PC acknowledge the financial support of the European Research Council (ERC; project PALs 320620). JDI gratefully acknowledges support from the DISCSIM project, grant agreement 341137, funded by the European Research Council under ERC-2013-ADG. ACB's contribution was supported, in part, by The University of British Columbia and the Canada Research Chairs program.

\appendix

\section{non-LTE Convergence}
\label{sec:convergence}

\begin{figure*}
    \centering
    \includegraphics[width=0.975\textwidth]{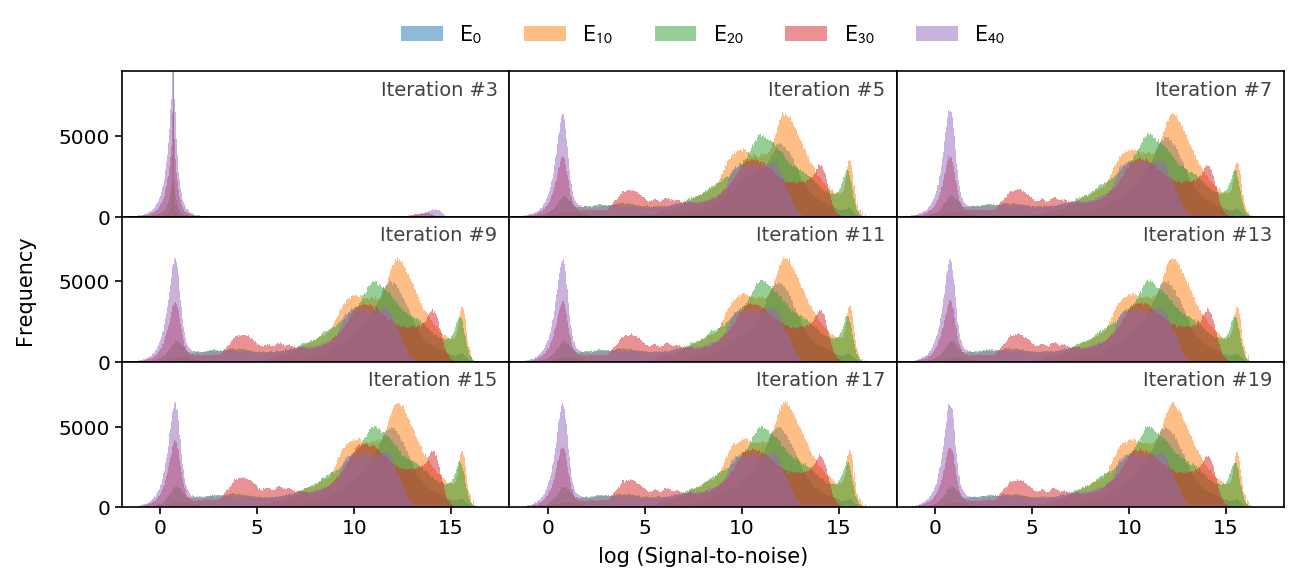}
    \caption{Histograms of the signal-to-noise ratios of different energy levels (i.e. $J$ number) for C$^{17}$O 1$\,\rightarrow\,$0 as the amount of iterations of level population calculations increases. A converged distribution is reached by 5 iterations. The non-LTE LIME run shown here uses 1 $\times$ 10$^6$ grid points and Sampling2.}
    \label{fig:snrs}
\end{figure*}

When non-LTE is adopted in LIME, there is no convergence threshold that, once surpassed, terminates the calculation of level populations. Instead, the user can set \textit{par$\rightarrow$nSolveIters} to dictate how many iterations will be performed. LIME does, however, calculate a signal-to-noise ratio for each grid point using the current iteration's level population and the standard deviation of the mean of the past five iterations' level populations. This information could be used to develop a convergence criterion, e.g. iterating stops once 90 per cent of grid points have reached a signal-to-noise ratio that exceeds 100, but this is beyond the scope of the current research. Instead, we explore the signal-to-noise ratio of the grid points as the number of iterations increases, and use this information to determine how many iterations we should use in our non-LTE runs to ensure a convergence level is reached. As \autoref{fig:snrs} indicates, this convergence level appears to be reached by five iterations, but in order to err on the side of caution we use ten iterations for our non-LTE runs.

\section{ALMA spectral resolution}

\begin{table}
    \centering
    \caption{Spectral resolution and channel spacing in kHz for different correlator bandwidth modes and channel averaging factor, $N$. Adapted from the ALMA Technical Handbook\protect\footnotemark.}
    
    \label{tab:almaspecres}
    \newcommand{\cspace}{\hspace*{5.0em}}
    \newcommand\B{\rule[-3.0ex]{0pt}{0pt}}
    \begin{tabular}{cC{1.25cm}cc}
        & & \cspace & \cspace \\ [-2.5ex]
        \toprule
        \multicolumn{1}{C{2.3cm}}{Usable bandwidth [MHz]} &  &  \multicolumn{2}{C{3.5cm}}{Spectral resolution (channel spacing) [kHz]} \B \\
        \cline{3-4}
        &  N =  &  1  &  2 \bigstrut[t] \\
        \midrule
        1875  &  &  977 (488)  &  1129 (977)  \\
        937.5  &  &  488 (244)  &  564  (488)  \\
        468.8  &  &  244 (122)  &  282  (244)  \\
        234.4  &  &  122 (61)  &  141  (122)  \\
        117.2  &  &  61 (31)  &  71  (61)  \\
        58.6  &  &  31 (15)  &  35  (31)  \\
        \bottomrule
        \end{tabular}
\end{table}

\footnotetext{https://almascience.eso.org/documents-and-tools/cycle5/alma-technical-handbook}

\autoref{tab:almaspecres} shows the spectral resolution attainable for ALMA for different usable bandwidths and channel averaging factors. Hanning smoothing is applied here, which results in a spectral resolution two times the channel spacing. As a result, it is recommended to increase the averaging factor, and therefore decrease the number of channels, because this results in a smaller loss in final resolution. However, in this paper we use $N$ = 1 in order to provide the most optimistic results, but note that this would require justification in an observational proposal.

\section{Flux dependencies}
\label{sec:fluxdependencies}

\begin{figure*}
    \centering
    \includegraphics[width=0.975\textwidth]{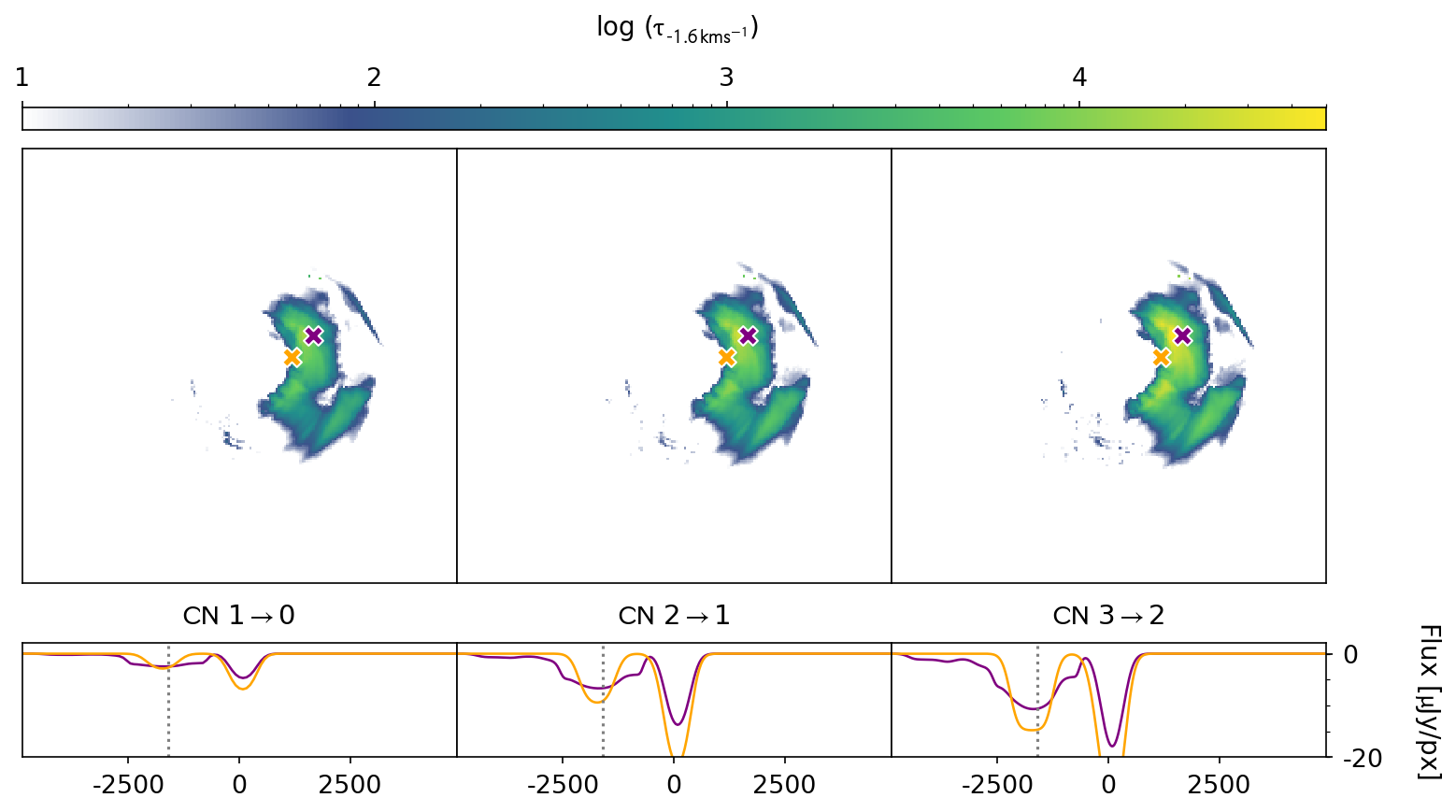}
    \caption{Comparison of CN absorption across different low-energy line transitions. The top panels show the optical depth at $v$ = -1.6$\,$km\,s$^{-1}$ for each molecule and the bottom panels show the line spectra at the pixel positions shown in the top panel, which is the same across all transitions. The dotted gray line on the bottom panels marks $v$ = 1.6$\,$km\,s$^{-1}$. These panels are produced for \lime\ images with no convolution applied.}
    \label{fig:CNComparisons}
\end{figure*}

\begin{figure*}
    \centering
    \includegraphics[width=0.975\textwidth]{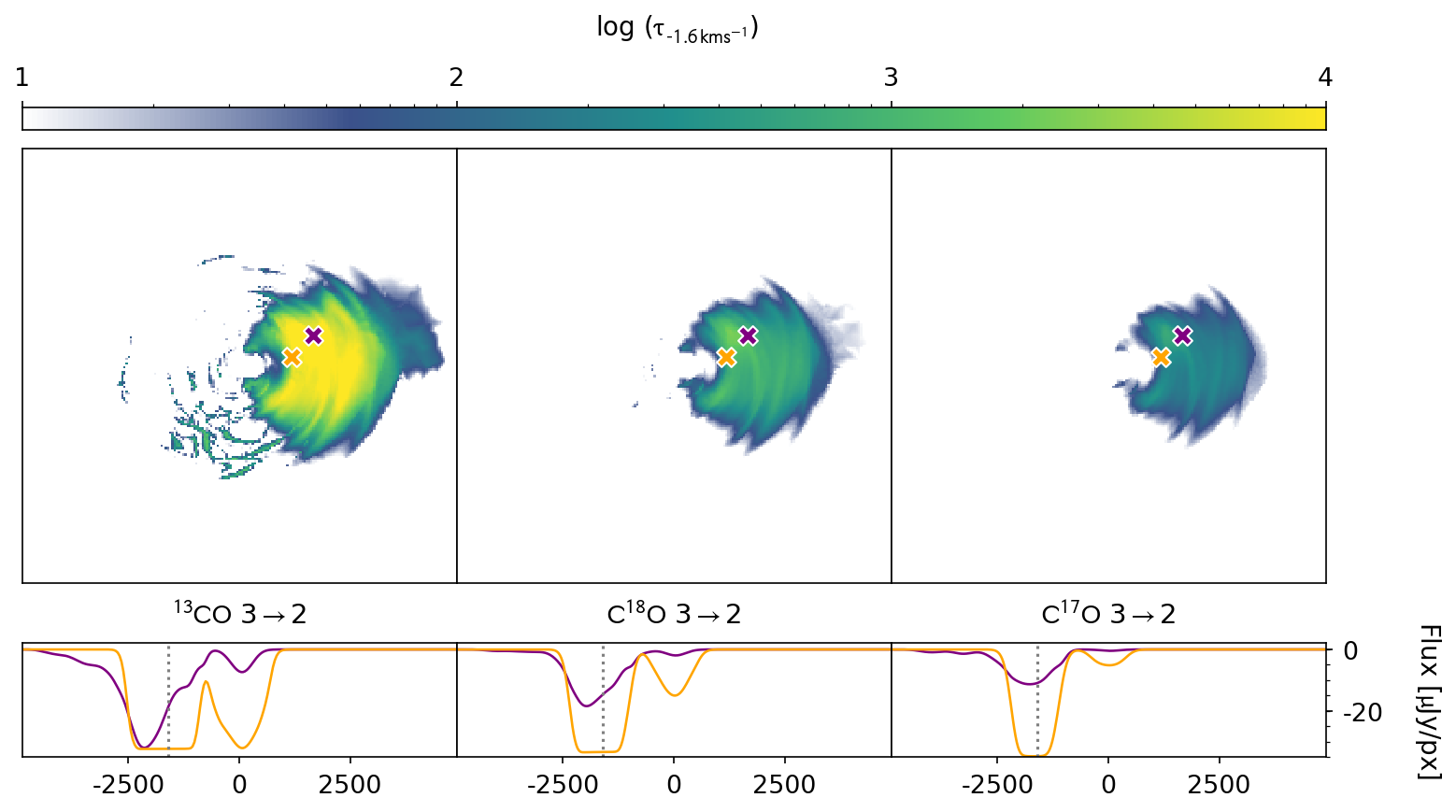}
    \caption{Same as \autoref{fig:CNComparisons} but for different CO isotopologues absorption.}
    \label{fig:COComparisons}
\end{figure*}

\autoref{fig:CNComparisons} shows the optical depth and example line spectra for different low energy transitions for the CN molecule, whereas \autoref{fig:COComparisons} shows the optical depth and example line spectra for different isotopologues of the CO molecule. Both figures together demonstrate that lower frequency transitions and lower abundance isotopologues are less optically thick; hence the absorption features are not as deep. At very high optical depths, certain transitions can become saturated. In this scenario, as is demonstrated by the orange line spectra in \autoref{fig:COComparisons}, the absorption feature widens rather than deepens with increased optical depth. 

\smallskip

We emphasise here that the line optical depths for the transitions shown in Figures \ref{fig:CNComparisons} and \ref{fig:COComparisons} are very large ($\tau$ \textgreater\ 100 at $v$ = -1.6\,km\,s$^{-1}$). Therefore, in young, embedded protoplanetary discs, even low frequency or low abundance isotopologues are likely to be optically thick. Consequently, only the surface layers of the disc are being probed, and properties derived from observations of Class 0 sources should account for this accordingly.


\begin{thebibliography}{}
\makeatletter
\relax
\def\mn@urlcharsother{\let\do\@makeother \do\$\do\&\do\#\do\^\do\_\do\%\do\~}
\def\mn@doi{\begingroup\mn@urlcharsother \@ifnextchar [ {\mn@doi@}
  {\mn@doi@[]}}
\def\mn@doi@[#1]#2{\def\@tempa{#1}\ifx\@tempa\@empty \href
  {http://dx.doi.org/#2} {doi:#2}\else \href {http://dx.doi.org/#2} {#1}\fi
  \endgroup}
\def\mn@eprint#1#2{\mn@eprint@#1:#2::\@nil}
\def\mn@eprint@arXiv#1{\href {http://arxiv.org/abs/#1} {{\tt arXiv:#1}}}
\def\mn@eprint@dblp#1{\href {http://dblp.uni-trier.de/rec/bibtex/#1.xml}
  {dblp:#1}}
\def\mn@eprint@#1:#2:#3:#4\@nil{\def\@tempa {#1}\def\@tempb {#2}\def\@tempc
  {#3}\ifx \@tempc \@empty \let \@tempc \@tempb \let \@tempb \@tempa \fi \ifx
  \@tempb \@empty \def\@tempb {arXiv}\fi \@ifundefined
  {mn@eprint@\@tempb}{\@tempb:\@tempc}{\expandafter \expandafter \csname
  mn@eprint@\@tempb\endcsname \expandafter{\@tempc}}}

\bibitem[\protect\citeauthoryear{{Ansdell}, {Williams}, {Manara}, {Miotello},
  {Facchini}, {van der Marel}, {Testi}  \& {van Dishoeck}}{{Ansdell}
  et~al.}{2017}]{Ansdell&Williams2017}
{Ansdell} M.,  {Williams} J.~P.,  {Manara} C.~F.,  {Miotello} A.,  {Facchini}
  S.,  {van der Marel} N.,  {Testi} L.,   {van Dishoeck} E.~F.,  2017, \mn@doi
  [AJ] {10.3847/1538-3881/aa69c0}, \href
  {http://adsabs.harvard.edu/abs/2017AJ....153..240A} {153, 240}

\bibitem[\protect\citeauthoryear{{Bae}, {Hartmann}, {Zhu}  \& {Nelson}}{{Bae}
  et~al.}{2014}]{Bae&Hartmann2014}
{Bae} J.,  {Hartmann} L.,  {Zhu} Z.,   {Nelson} R.~P.,  2014, \mn@doi [ApJ]
  {10.1088/0004-637X/795/1/61}, \href
  {http://adsabs.harvard.edu/abs/2014ApJ...795...61B} {795, 61}

\bibitem[\protect\citeauthoryear{{Bissantz}, {Englmaier}  \&
  {Gerhard}}{{Bissantz} et~al.}{2003}]{Bissantz&Englmaier2003}
{Bissantz} N.,  {Englmaier} P.,   {Gerhard} O.,  2003, \mn@doi [MNRAS]
  {10.1046/j.1365-8711.2003.06358.x}, \href
  {http://adsabs.harvard.edu/abs/2003MNRAS.340..949B} {340, 949}

\bibitem[\protect\citeauthoryear{{Boley} \& {Durisen}}{{Boley} \&
  {Durisen}}{2008}]{Boley&Durisen2008}
{Boley} A.~C.,  {Durisen} R.~H.,  2008, \mn@doi [ApJ] {10.1086/591013}, \href
  {http://adsabs.harvard.edu/abs/2008ApJ...685.1193B} {685, 1193}

\bibitem[\protect\citeauthoryear{{Brinch} \& {Hogerheijde}}{{Brinch} \&
  {Hogerheijde}}{2010}]{Brinch&Hogerheijde2010}
{Brinch} C.,  {Hogerheijde} M.~R.,  2010, \mn@doi [A\&A]
  {10.1051/0004-6361/201015333}, \href
  {http://adsabs.harvard.edu/abs/2010A%26A...523A..25B} {523, A25}

\bibitem[\protect\citeauthoryear{{Chapillon}, {Guilloteau}, {Dutrey},
  {Pi{\'e}tu}  \& {Gu{\'e}lin}}{{Chapillon}
  et~al.}{2012}]{Chapillon&Guilloteau2012}
{Chapillon} E.,  {Guilloteau} S.,  {Dutrey} A.,  {Pi{\'e}tu} V.,   {Gu{\'e}lin}
  M.,  2012, \mn@doi [A\&A] {10.1051/0004-6361/201116762}, \href
  {http://adsabs.harvard.edu/abs/2012A%26A...537A..60C} {537, A60}

\bibitem[\protect\citeauthoryear{{Cossins}, {Lodato}  \& {Testi}}{{Cossins}
  et~al.}{2010}]{Cossins&Lodato2010}
{Cossins} P.,  {Lodato} G.,   {Testi} L.,  2010, \mn@doi [MNRAS]
  {10.1111/j.1365-2966.2010.16934.x}, \href
  {http://adsabs.harvard.edu/abs/2010MNRAS.407..181C} {407, 181}

\bibitem[\protect\citeauthoryear{{D'Alessio}, {Calvet}, {Hartmann},
  {Franco-Hern{\'a}ndez}  \& {Serv{\'{\i}}n}}{{D'Alessio}
  et~al.}{2006}]{DAlessio&Calvet2006}
{D'Alessio} P.,  {Calvet} N.,  {Hartmann} L.,  {Franco-Hern{\'a}ndez} R.,
  {Serv{\'{\i}}n} H.,  2006, \mn@doi [ApJ] {10.1086/498861}, \href
  {http://adsabs.harvard.edu/abs/2006ApJ...638..314D} {638, 314}

\bibitem[\protect\citeauthoryear{{Dipierro}, {Lodato}, {Testi}  \& {de Gregorio
  Monsalvo}}{{Dipierro} et~al.}{2014}]{Dipierro&Lodato2014}
{Dipierro} G.,  {Lodato} G.,  {Testi} L.,   {de Gregorio Monsalvo} I.,  2014,
  \mn@doi [MNRAS] {10.1093/mnras/stu1584}, \href
  {http://adsabs.harvard.edu/abs/2014MNRAS.444.1919D} {444, 1919}

\bibitem[\protect\citeauthoryear{{Dipierro}, {Pinilla}, {Lodato}  \&
  {Testi}}{{Dipierro} et~al.}{2015}]{Dipierro&Pinilla2015}
{Dipierro} G.,  {Pinilla} P.,  {Lodato} G.,   {Testi} L.,  2015, \mn@doi
  [MNRAS] {10.1093/mnras/stv970}, \href
  {https://ui.adsabs.harvard.edu/#abs/2015MNRAS.451..974D} {451, 974}

\bibitem[\protect\citeauthoryear{{Douglas}, {Caselli}, {Ilee}, {Boley},
  {Hartquist}, {Durisen}  \& {Rawlings}}{{Douglas}
  et~al.}{2013}]{Douglas&Caselli2013}
{Douglas} T.~A.,  {Caselli} P.,  {Ilee} J.~D.,  {Boley} A.~C.,  {Hartquist}
  T.~W.,  {Durisen} R.~H.,   {Rawlings} J.~M.~C.,  2013, \mn@doi [MNRAS]
  {10.1093/mnras/stt881}, \href
  {http://adsabs.harvard.edu/abs/2013MNRAS.433.2064D} {433, 2064}

\bibitem[\protect\citeauthoryear{{Drozdovskaya}, {Walsh}, {Visser}, {Harsono}
  \& {van Dishoeck}}{{Drozdovskaya} et~al.}{2015}]{Drozdovskaya&Walsh2015}
{Drozdovskaya} M.~N.,  {Walsh} C.,  {Visser} R.,  {Harsono} D.,   {van
  Dishoeck} E.~F.,  2015, \mn@doi [MNRAS] {10.1093/mnras/stv1177}, \href
  {http://adsabs.harvard.edu/abs/2015MNRAS.451.3836D} {451, 3836}

\bibitem[\protect\citeauthoryear{{Durisen}, {Boss}, {Mayer}, {Nelson}, {Quinn}
  \& {Rice}}{{Durisen} et~al.}{2007}]{Durisen&Boss2007}
{Durisen} R.~H.,  {Boss} A.~P.,  {Mayer} L.,  {Nelson} A.~F.,  {Quinn} T.,
  {Rice} W.~K.~M.,  2007, in Protostars and Planets V. p.~607 (\mn@eprint
  {arXiv} {astro-ph/0603179})

\bibitem[\protect\citeauthoryear{{Dutrey}, {Guilloteau}  \& {Guelin}}{{Dutrey}
  et~al.}{1997}]{Dutrey&Guilloteau1997}
{Dutrey} A.,  {Guilloteau} S.,   {Guelin} M.,  1997, A\&A, \href
  {http://adsabs.harvard.edu/abs/1997A%26A...317L..55D} {317, L55}

\bibitem[\protect\citeauthoryear{{Dutrey} et~al.,}{{Dutrey}
  et~al.}{2011}]{Dutrey&Wakelam2011}
{Dutrey} A.,  et~al., 2011, \mn@doi [A\&A] {10.1051/0004-6361/201116931}, \href
  {http://adsabs.harvard.edu/abs/2011A%26A...535A.104D} {535, A104}

\bibitem[\protect\citeauthoryear{{Evans}, {Ilee}, {Boley}, {Caselli},
  {Durisen}, {Hartquist}  \& {Rawlings}}{{Evans} et~al.}{2015}]{Evans&Ilee2015}
{Evans} M.~G.,  {Ilee} J.~D.,  {Boley} A.~C.,  {Caselli} P.,  {Durisen} R.~H.,
  {Hartquist} T.~W.,   {Rawlings} J.~M.~C.,  2015, \mn@doi [MNRAS]
  {10.1093/mnras/stv1698}, \href
  {http://adsabs.harvard.edu/abs/2015MNRAS.453.1147E} {453, 1147}

\bibitem[\protect\citeauthoryear{{Evans} et~al.,}{{Evans}
  et~al.}{2017}]{Evans&Ilee2017}
{Evans} M.~G.,  et~al., 2017, \mn@doi [MNRAS] {10.1093/mnras/stx1365}, \href
  {http://adsabs.harvard.edu/abs/2017MNRAS.470.1828E} {470, 1828}

\bibitem[\protect\citeauthoryear{{Gammie}}{{Gammie}}{2001}]{Gammie2001}
{Gammie} C.~F.,  2001, \mn@doi [ApJ] {10.1086/320631}, \href
  {https://ui.adsabs.harvard.edu/#abs/2001ApJ...553..174G} {553, 174}

\bibitem[\protect\citeauthoryear{{Gerin} et~al.,}{{Gerin}
  et~al.}{2017}]{Gerin&Pety2017}
{Gerin} M.,  et~al., 2017, \mn@doi [A\&A] {10.1051/0004-6361/201630187}, \href
  {http://adsabs.harvard.edu/abs/2017A%26A...606A..35G} {606, A35}

\bibitem[\protect\citeauthoryear{{Guilloteau} et~al.,}{{Guilloteau}
  et~al.}{2016}]{Guilloteau&Reboussin2016}
{Guilloteau} S.,  et~al., 2016, \mn@doi [A\&A] {10.1051/0004-6361/201527088},
  \href {http://adsabs.harvard.edu/abs/2016A%26A...592A.124G} {592, A124}

\bibitem[\protect\citeauthoryear{{Hall}, {Forgan}, {Rice}, {Harries},
  {Klaassen}  \& {Biller}}{{Hall} et~al.}{2016}]{Hall&Forgan2016}
{Hall} C.,  {Forgan} D.,  {Rice} K.,  {Harries} T.~J.,  {Klaassen} P.~D.,
  {Biller} B.,  2016, \mn@doi [MNRAS] {10.1093/mnras/stw296}, \href
  {http://adsabs.harvard.edu/abs/2016MNRAS.458..306H} {458, 306}

\bibitem[\protect\citeauthoryear{{Harker} \& {Desch}}{{Harker} \&
  {Desch}}{2002}]{Harker&Desch2002}
{Harker} D.~E.,  {Desch} S.~J.,  2002, \mn@doi [ApJ] {10.1086/339363}, \href
  {http://adsabs.harvard.edu/abs/2002ApJ...565L.109H} {565, L109}

\bibitem[\protect\citeauthoryear{{Harsono}, {Alexander}  \& {Levin}}{{Harsono}
  et~al.}{2011}]{Harsono&Alexander2011}
{Harsono} D.,  {Alexander} R.~D.,   {Levin} Y.,  2011, \mn@doi [MNRAS]
  {10.1111/j.1365-2966.2010.18146.x}, \href
  {https://ui.adsabs.harvard.edu/#abs/2011MNRAS.413..423H} {413, 423}

\bibitem[\protect\citeauthoryear{Henning \& Semenov}{Henning \&
  Semenov}{2013}]{Henning&Semenov2013}
Henning T.,  Semenov D.,  2013, \mn@doi [Chemical Reviews] {10.1021/cr400128p},
  \href {http://adsabs.harvard.edu/abs/2013ChRv..113.9016H} {113, 9016}

\bibitem[\protect\citeauthoryear{{Ilee}, {Boley}, {Caselli}, {Durisen},
  {Hartquist}  \& {Rawlings}}{{Ilee} et~al.}{2011}]{Ilee&Caselli2011}
{Ilee} J.~D.,  {Boley} A.~C.,  {Caselli} P.,  {Durisen} R.~H.,  {Hartquist}
  T.~W.,   {Rawlings} J.~M.~C.,  2011, \mn@doi [MNRAS]
  {10.1111/j.1365-2966.2011.19455.x}, \href
  {http://adsabs.harvard.edu/abs/2011MNRAS.417.2950I} {417, 2950}

\bibitem[\protect\citeauthoryear{{Kama} et~al.,}{{Kama}
  et~al.}{2016a}]{Kama&Bruderer2016a}
{Kama} M.,  et~al., 2016a, \mn@doi [A\&A] {10.1051/0004-6361/201526791}, \href
  {https://ui.adsabs.harvard.edu/#abs/2016A&A...588A.108K} {588}

\bibitem[\protect\citeauthoryear{{Kama} et~al.,}{{Kama}
  et~al.}{2016b}]{Kama&Bruderer2016b}
{Kama} M.,  et~al., 2016b, \mn@doi [A\&A] {10.1051/0004-6361/201526991}, \href
  {https://ui.adsabs.harvard.edu/#abs/2016A&A...592A..83K} {592}

\bibitem[\protect\citeauthoryear{{Keto}, {Caselli}  \& {Rawlings}}{{Keto}
  et~al.}{2015}]{Keto&Caselli2015}
{Keto} E.,  {Caselli} P.,   {Rawlings} J.,  2015, \mn@doi [MNRAS]
  {10.1093/mnras/stu2247}, \href
  {http://adsabs.harvard.edu/abs/2015MNRAS.446.3731K} {446, 3731}

\bibitem[\protect\citeauthoryear{{Koumpia}, {Semenov}, {van der Tak}, {Boogert}
   \& {Caux}}{{Koumpia} et~al.}{2017}]{Koumpia&Semenov2017}
{Koumpia} E.,  {Semenov} D.~A.,  {van der Tak} F.~F.~S.,  {Boogert} A.~C.~A.,
  {Caux} E.,  2017, \mn@doi [A\&A] {10.1051/0004-6361/201630160}, \href
  {http://adsabs.harvard.edu/abs/2017A%26A...603A..88K} {603, A88}

\bibitem[\protect\citeauthoryear{{Li}, {Gerhard}, {Shen}, {Portail}  \&
  {Wegg}}{{Li} et~al.}{2016}]{Li&Gerhard2016}
{Li} Z.,  {Gerhard} O.,  {Shen} J.,  {Portail} M.,   {Wegg} C.,  2016, \mn@doi
  [ApJ] {10.3847/0004-637X/824/1/13}, \href
  {http://adsabs.harvard.edu/abs/2016ApJ...824...13L} {824, 13}

\bibitem[\protect\citeauthoryear{{Lodato} \& {Rice}}{{Lodato} \&
  {Rice}}{2004}]{Lodato&Rice2004}
{Lodato} G.,  {Rice} W.~K.~M.,  2004, \mn@doi [MNRAS]
  {10.1111/j.1365-2966.2004.07811.x}, \href
  {https://ui.adsabs.harvard.edu/#abs/2004MNRAS.351..630L} {351, 630}

\bibitem[\protect\citeauthoryear{{Machida} \& {Matsumoto}}{{Machida} \&
  {Matsumoto}}{2011}]{Machida&Matsumoto2011}
{Machida} M.~N.,  {Matsumoto} T.,  2011, \mn@doi [MNRAS]
  {10.1111/j.1365-2966.2011.18349.x}, \href
  {http://adsabs.harvard.edu/abs/2011MNRAS.413.2767M} {413, 2767}

\bibitem[\protect\citeauthoryear{{McMullin}, {Waters}, {Schiebel}, {Young}  \&
  {Golap}}{{McMullin} et~al.}{2007}]{McMullin&Waters2007}
{McMullin} J.~P.,  {Waters} B.,  {Schiebel} D.,  {Young} W.,   {Golap} K.,
  2007, in {Shaw} R.~A.,  {Hill} F.,   {Bell} D.~J.,  eds,  Astronomical
  Society of the Pacific Conference Series Vol. 376, Astronomical Data Analysis
  Software and Systems XVI. p.~127

\bibitem[\protect\citeauthoryear{{Meru}, {Juh{\'a}sz}, {Ilee}, {Clarke},
  {Rosotti}  \& {Booth}}{{Meru} et~al.}{2017}]{Meru&Juhasz2017}
{Meru} F.,  {Juh{\'a}sz} A.,  {Ilee} J.~D.,  {Clarke} C.~J.,  {Rosotti} G.~P.,
   {Booth} R.~A.,  2017, \mn@doi [ApJ] {10.3847/2041-8213/aa6837}, \href
  {https://ui.adsabs.harvard.edu/#abs/2017ApJ...839L..24M} {839}

\bibitem[\protect\citeauthoryear{{Miettinen}}{{Miettinen}}{2016}]{Miettinen2016}
{Miettinen} O.,  2016, \mn@doi [Ap\&SS] {10.1007/s10509-016-2834-9}, \href
  {http://adsabs.harvard.edu/abs/2016Ap%26SS.361..248M} {361, 248}

\bibitem[\protect\citeauthoryear{{Miotello} et~al.,}{{Miotello}
  et~al.}{2017}]{Miotello&vanDishoeck2017}
{Miotello} A.,  et~al., 2017, \mn@doi [A\&A] {10.1051/0004-6361/201629556},
  \href {http://adsabs.harvard.edu/abs/2017A%26A...599A.113M} {599, A113}

\bibitem[\protect\citeauthoryear{{{\"O}berg}, {Guzm{\'a}n}, {Furuya}, {Qi},
  {Aikawa}, {Andrews}, {Loomis}  \& {Wilner}}{{{\"O}berg}
  et~al.}{2015}]{Oberg&Guzman2015}
{{\"O}berg} K.~I.,  {Guzm{\'a}n} V.~V.,  {Furuya} K.,  {Qi} C.,  {Aikawa} Y.,
  {Andrews} S.~M.,  {Loomis} R.,   {Wilner} D.~J.,  2015, \mn@doi [Nature]
  {10.1038/nature14276}, \href
  {http://adsabs.harvard.edu/abs/2015Natur.520..198O} {520, 198}

\bibitem[\protect\citeauthoryear{{Ohashi} et~al.,}{{Ohashi}
  et~al.}{2014}]{Ohashi&Saigo2014}
{Ohashi} N.,  et~al., 2014, \mn@doi [ApJ] {10.1088/0004-637X/796/2/131}, \href
  {http://adsabs.harvard.edu/abs/2014ApJ...796..131O} {796, 131}

\bibitem[\protect\citeauthoryear{{P{\'e}rez} et~al.,}{{P{\'e}rez}
  et~al.}{2016}]{Perez&Carpenter2016}
{P{\'e}rez} L.~M.,  et~al., 2016, \mn@doi [Science] {10.1126/science.aaf8296},
  \href {http://adsabs.harvard.edu/abs/2016Sci...353.1519P} {353, 1519}

\bibitem[\protect\citeauthoryear{{Qi}, {{\"O}berg}, {Wilner}  \&
  {Rosenfeld}}{{Qi} et~al.}{2013}]{Qi&Oberg2013}
{Qi} C.,  {{\"O}berg} K.~I.,  {Wilner} D.~J.,   {Rosenfeld} K.~A.,  2013,
  \mn@doi [ApJ] {10.1088/2041-8205/765/1/L14}, \href
  {http://adsabs.harvard.edu/abs/2013ApJ...765L..14Q} {765, L14}

\bibitem[\protect\citeauthoryear{{Rodriguez-Fernandez} \&
  {Combes}}{{Rodriguez-Fernandez} \& {Combes}}{2008}]{Rodriguez&Combes2008}
{Rodriguez-Fernandez} N.~J.,  {Combes} F.,  2008, \mn@doi [A\&A]
  {10.1051/0004-6361:200809644}, \href
  {http://adsabs.harvard.edu/abs/2008A%26A...489..115R} {489, 115}

\bibitem[\protect\citeauthoryear{Sakai et~al.,}{Sakai
  et~al.}{2014a}]{Sakai&Sakai2014}
Sakai N.,  et~al., 2014a, \mn@doi [Nature] {10.1038/nature13000}, \href
  {http://adsabs.harvard.edu/abs/2014Natur.507...78S} {507, 78}

\bibitem[\protect\citeauthoryear{{Sakai} et~al.,}{{Sakai}
  et~al.}{2014b}]{Sakai&Oya2014}
{Sakai} N.,  et~al., 2014b, \mn@doi [ApJ] {10.1088/2041-8205/791/2/L38}, \href
  {http://adsabs.harvard.edu/abs/2014ApJ...791L..38S} {791, L38}

\bibitem[\protect\citeauthoryear{{Sakai} et~al.,}{{Sakai}
  et~al.}{2016}]{Sakai&Oya2016}
{Sakai} N.,  et~al., 2016, \mn@doi [ApJ] {10.3847/2041-8205/820/2/L34}, \href
  {http://adsabs.harvard.edu/abs/2016ApJ...820L..34S} {820, L34}

\bibitem[\protect\citeauthoryear{{Sakai} et~al.,}{{Sakai}
  et~al.}{2017}]{Sakai&Oya2017}
{Sakai} N.,  et~al., 2017, \mn@doi [MNRAS] {10.1093/mnrasl/slx002}, \href
  {http://adsabs.harvard.edu/abs/2017MNRAS.467L..76S} {467, L76}

\bibitem[\protect\citeauthoryear{{Salinas} et~al.,}{{Salinas}
  et~al.}{2016}]{Salinas&Hogerheijde2016}
{Salinas} V.~N.,  et~al., 2016, \mn@doi [A\&A] {10.1051/0004-6361/201628172},
  \href {http://adsabs.harvard.edu/abs/2016A%26A...591A.122S} {591, A122}

\bibitem[\protect\citeauthoryear{{Salinas}, {Hogerheijde}, {Mathews},
  {{\"O}berg}, {Qi}, {Williams}  \& {Wilner}}{{Salinas}
  et~al.}{2017}]{Salinas&Hogerheijde2017}
{Salinas} V.~N.,  {Hogerheijde} M.~R.,  {Mathews} G.~S.,  {{\"O}berg} K.~I.,
  {Qi} C.,  {Williams} J.~P.,   {Wilner} D.~J.,  2017, \mn@doi [A\&A]
  {10.1051/0004-6361/201731223}, \href
  {http://adsabs.harvard.edu/abs/2017A%26A...606A.125S} {606, A125}

\bibitem[\protect\citeauthoryear{{Seifried}, {S{\'a}nchez-Monge}, {Walch}  \&
  {Banerjee}}{{Seifried} et~al.}{2016}]{Seifried&Sanchez-Monge2016}
{Seifried} D.,  {S{\'a}nchez-Monge} {\'A}.,  {Walch} S.,   {Banerjee} R.,
  2016, \mn@doi [MNRAS] {10.1093/mnras/stw785}, \href
  {http://adsabs.harvard.edu/abs/2016MNRAS.459.1892S} {459, 1892}

\bibitem[\protect\citeauthoryear{{Shirley}}{{Shirley}}{2015}]{Shirley2015}
{Shirley} Y.~L.,  2015, \mn@doi [PASP] {10.1086/680342}, \href
  {http://adsabs.harvard.edu/abs/2015PASP..127..299S} {127, 299}

\bibitem[\protect\citeauthoryear{{Thi} et~al.,}{{Thi}
  et~al.}{2011}]{Thi&Menard2011}
{Thi} W.-F.,  et~al., 2011, \mn@doi [A\&A] {10.1051/0004-6361/201116678}, \href
  {http://adsabs.harvard.edu/abs/2011A%26A...530L...2T} {530, L2}

\bibitem[\protect\citeauthoryear{{Tobin}, {Hartmann}, {Chiang}, {Wilner},
  {Looney}, {Loinard}, {Calvet}  \& {D'Alessio}}{{Tobin}
  et~al.}{2012}]{Tobin&Hartmann2012}
{Tobin} J.~J.,  {Hartmann} L.,  {Chiang} H.-F.,  {Wilner} D.~J.,  {Looney}
  L.~W.,  {Loinard} L.,  {Calvet} N.,   {D'Alessio} P.,  2012, \mn@doi [Nature]
  {10.1038/nature11610}, \href
  {http://adsabs.harvard.edu/abs/2012Natur.492...83T} {492, 83}

\bibitem[\protect\citeauthoryear{{Tobin} et~al.,}{{Tobin}
  et~al.}{2016}]{Tobin&Kratter2016}
{Tobin} J.~J.,  et~al., 2016, \mn@doi [Nature] {10.1038/nature20094}, \href
  {http://adsabs.harvard.edu/abs/2016Natur.538..483T} {538, 483}

\bibitem[\protect\citeauthoryear{{Walsh} et~al.,}{{Walsh}
  et~al.}{2016}]{Walsh&Loomis2016}
{Walsh} C.,  et~al., 2016, \mn@doi [ApJ] {10.3847/2041-8205/823/1/L10}, \href
  {http://adsabs.harvard.edu/abs/2016ApJ...823L..10W} {823, L10}

\bibitem[\protect\citeauthoryear{{Williams} \& {Cieza}}{{Williams} \&
  {Cieza}}{2011}]{Williams&Cieza2011}
{Williams} J.~P.,  {Cieza} L.~A.,  2011, \mn@doi [ARA\&A]
  {10.1146/annurev-astro-081710-102548}, \href
  {http://adsabs.harvard.edu/abs/2011ARA%26A..49...67W} {49, 67}

\bibitem[\protect\citeauthoryear{{Wilson}}{{Wilson}}{1999}]{Wilson1999}
{Wilson} T.~L.,  1999, \mn@doi [Reports on Progress in Physics]
  {10.1088/0034-4885/62/2/002}, \href
  {http://adsabs.harvard.edu/abs/1999RPPh...62..143W} {62, 143}

\bibitem[\protect\citeauthoryear{{Young}, {Lee}, {Evans}, {Goldsmith}  \&
  {Doty}}{{Young} et~al.}{2004}]{Young&Lee2004}
{Young} K.~E.,  {Lee} J.-E.,  {Evans} II N.~J.,  {Goldsmith} P.~F.,   {Doty}
  S.~D.,  2004, \mn@doi [ApJ] {10.1086/423609}, \href
  {http://adsabs.harvard.edu/abs/2004ApJ...614..252Y} {614, 252}

\bibitem[\protect\citeauthoryear{{Yu}, {Evans}, {Dodson-Robinson}, {Willacy}
  \& {Turner}}{{Yu} et~al.}{2017}]{Yu&Evans2017}
{Yu} M.,  {Evans} II N.~J.,  {Dodson-Robinson} S.~E.,  {Willacy} K.,   {Turner}
  N.~J.,  2017, \mn@doi [ApJ] {10.3847/1538-4357/aa6e4c}, \href
  {http://adsabs.harvard.edu/abs/2017ApJ...841...39Y} {841, 39}

\makeatother
\end{thebibliography}
\end{document}